\documentclass[preprint2,times,tighten]{aastex6}
\usepackage{amsmath,amstext}
\usepackage[all]{hypcap} 
\usepackage[english]{babel}
\usepackage[utf8x]{inputenc}
\usepackage{pgfplots}
\usepackage{graphicx}
\pgfplotsset{compat=newest}
\usepackage{epsfig}
\usepackage{color}
\usepackage[hang]{subfigure}
 
\newcommand{\teff}{$T_\mathrm{eff}$}
\newcommand{\logg}{$\log g$}
\newcommand{\feh}{[Fe/H]}

\newcommand{\kms}{km\,s$^{-1}$}
\newcommand{\mic}{$\mu \mathrm m$}

\shorttitle{Chemistry of the Galactic Center}
\shortauthors{Rich et al.}

\begin{document}

\title{Detailed Abundances for the Old Population near the Galactic Center:\\
I. Metallicity  distribution of the Nuclear Star Cluster}
\author{R. M. Rich\altaffilmark{1}, N. Ryde\altaffilmark{2}, B. Thorsbro\altaffilmark{2}, T. K. Fritz\altaffilmark{3}, M. Schultheis\altaffilmark{4}, L. Origlia\altaffilmark{5}, H. J\"onsson\altaffilmark{2}
}
 \email{ryde@astro.lu.se}
 
\altaffiltext{1}{Department of Physics and Astronomy, UCLA, 430 Portola Plaza, Box 951547, Los Angeles, CA 90095-1547, USA}
\altaffiltext{2}{Lund Observatory, Department of Astronomy and Theoretical Physics, Lund University, Box 43, SE-221 00 Lund, Sweden}
\altaffiltext{3}{Department of Astronomy, University of Virginia, 3530 McCormick Road, Charlottesville, VA 22904, USA}
\altaffiltext{4}{Observatoire de la C\^ote d'Azur, CNRS UMR 7293, BP4229, Laboratoire Lagrange, F-06304 Nice Cedex 4, France}
\altaffiltext{5}{INAF - Osservatorio Astronomico di Bologna, Via Gobetti 93/3, I-40129 Bologna, Italy}

\begin{abstract}
We report the first high spectral resolution study of 17 M giants kinematically confirmed to lie within a few parsecs of the Galactic Center, using $R\sim 24,000$ spectroscopy from Keck/NIRSPEC and a new linelist for the infrared K band.  We consider their luminosities and kinematics, which classify these stars as members of the older stellar population and the central cluster.  We find a median metallicity of $<$\feh$>=-0.16$ and a large spread from approximately $-0.3$ to $+0.3$ (quartiles).  We find that the highest metallicities are $\rm [Fe/H]<+0.6$, with most of the stars being at or below the Solar iron abundance.  The abundances and the abundance distribution strongly resembles that of the Galactic bulge rather than disk or halo; in our small sample we find no statistical evidence for a dependence of velocity dispersion on metallicity.  


\end{abstract}

\maketitle

\section{Introduction}
Only in recent years has it become feasible to explore the composition of stars that are members of the central cluster that occupies the inner parsec of the Milky Way.  This region continues to be of great interest as most of its mass resides within the sphere of influence of the supermassive black hole, Sgr A*.  The stellar population surrounding this cluster is the bar/bulge system, which now has $\approx 10^4$ members with high resolution and high S/N optical/IR spectra.  Nonetheless, one cannot assume that there is any connection between the central cluster and the bar/bulge population as a whole.  While the bar/bulge population is dominated by an older stellar population  \citep[e.g.][]{kuijken:02,clarkson} the central cluster still has ongoing star formation \citep[e.g.][]{Paumard:06,Feldmeier-Krause:15}.  And microlensing studies \citep[e.g.][]{bensby:13,bensby:17} argue for the presence of a metal rich, intermediate age population in the bulge.  Separated from the bar/bulge by its dense structure and kinematics, its potential well defined by both Sgr A* and the stellar cluster, the central cluster might well be expected to have experienced a different and unique history of chemical evolution, leaving its own imprint of chemical abundances.  This is especially true because of the presence of Sgr A*, the evidence for a complex star formation history \citep{Pfuhl_11} suggesting cross pollination with gas from the disk and inner bulge regions. Of additional interest is the possibility of gaining insight into the early chemical evolution of this special environment. The presence of stars older than 1 Gyr (and likely 8-10 Gyr old)  within 50 pc of the Galactic center has been established since \citet{figer} clearly detected the red clump, and the discovery of RR Lyrae variables near Sgr A* \citep{dong:17} argues for at least a significant fraction of the central cluster being globular cluster age i.e. $\sim 10$ Gyr old.  The expected compositions at this Galactic crossroads might be alpha-enhanced from an early starburst, or more toward scaled Solar, reflecting the extended star formation history of this region.  The reality is that we will probably find representatives of both ancient and relatively recent star formation; more complete inferences must await larger samples with analyses of additional elements.

Spectroscopy in the central parsec has focused on either the brightest stars, that are likely supergiants, or on fainter red giants that are likely to be dominated in number by the $\sim 10$ Gyr old population because lifetimes of those stars on the red giant branch are longer.  Early studies emphasized the brighter red supergiants, with all stars $K<9.5$ and $M_\mathrm{bol}<-5$, with the best results from IRS7, the most luminous red supergiant \citep{ramirez:00a,cunha:2007:apj}.  These studies found a Solar iron abundance  with alpha enhancement for this population.  High resolution spectroscopy of nine fainter red giants by  \citet{ryde_schultheis:15} find a suprasolar distribution in [Fe/H] but otherwise normal $[\alpha/\rm Fe]$.   

 Medium resolution spectroscopy has enabled AO-assisted \citep{do:15} and wider field \citep{Feldmeier-Krause:17} studies of fainter red giants in the inner parsec (Do et al. 2015) or near the Galactic center, at spectral resolutions of $\sim 4000-5000$.  The medium resolution studies have found evidence for a metal poor population that comprises roughly 10\% of the stellar population.  The metal poor, alpha enhanced nature of one kinematically confirmed Galactic Center red giant has been determined by our team using high resolution spectroscopy \citep{Ryde:GC:MPG}, but the metallicity distribution of the metal rich stellar component in the Galactic Center remains a subject of debate.  \citet{rich:12} examined high resolution ($R=23,000$) H-band spectra of 58 red giants ranging from 140-400 pc from the nucleus, finding no abundance gradient and $\langle \rm [Fe/H] \rangle=-0.2$.  The adaptive optics-aided investigation of the central parsec by  \citet{do:15} found the majority of stars with supra-solar metallicities, including one star with [Fe/H]=+0.96.  \citet{Feldmeier-Krause:17} used a Bayesian approach to derive stellar parameters and abundances similar to that of \citet{do:15} and found $\langle \rm  [Fe/H] \rangle =+0.26$ with stars reaching nearly 10 times Solar as well.  While the inferred surface gravities of \citet{do:15} were of concern, sometimes reaching \logg$\approx 4$ for red giants, those of \citet{Feldmeier-Krause:17} are more in line with stellar parameters expected for an old stellar population at 8 kpc.  Nonetheless, it is a truly daunting task to undertake abundance analysis including derivation of stellar parameters, for relatively low S/N spectra at $R\sim 4-5000$, especially for metal-rich, cool stars.  The ubiquitous blending and lack of continuum points legitimately raises the question as to whether any analysis at such low resolution is even possible.  When strong lines can be identified in the spectra, they are certain to be so strongly saturated and blended that no trustworthy derivation of abundance by traditional means is possible. In general, abundance determinations at different resolutions are a complex interplay of accuracy,  metallicity, effective temperature, spectral band and coverage, as well as signal-to-noise ratio. 

The need to establish the metallicity distribution with precision has heightened as new studies of the Galactic Center question the origin of the population.  \citet{minniti:16} report RR Lyrae in the "nuclear" bulge region, within 1$^o$ of the Galactic Center.  They propose that the presence of RR Lyrae imply that the nucleus was built from the debris of globular clusters and therefore from an old metal-poor population.  The confirmation of RR Lyrae near Sgr A* potentially extends Minniti's model of globular cluster accretion to the nucleus itself; our measurement of the abundance distribution can test this hypothesis.

Here we report [Fe/H] for giants in the Galactic Center from high resolution $(R\sim24000)$ spectra in the $2.4\,\mu$m K band, obtained using NIRSPEC at the W.M. Keck Observatory.  We have spectra covering most of the K band for stars with $12 <K < 10.5$. We have analyzed our spectra using a high resolution detailed abundance analysis, with the benefit of a new linelist (Thorsbro et al. 2018). The current paper is the first in a series, in which forthcoming papers will deal with specific elemental abundances, such as those of the $\alpha$ elements and that of  Sc, and V, as well as comparative studies to other Bulge fields.

\section{OBSERVATIONS}

We have obtained high resolution ($R\sim 24,000$) infrared K band spectra for18 M-giants belonging to the Nuclear Star Cluster (NSC) of the Milky Way, using NIRSPEC \citep{nirspec,nirspec_mclean} at Keck II. The observations were done during four half-nights on April, 27-28 2015 and April 18-19 2016.  In Table \ref{tab:star} their coordinates and total exposure times are given.  The  galactic coordinates in the Table are seemingly offset from Sgr A*, because the center of the galactic coordinate system was defined in 1958 \citep{blaauw:60} and is offset relative to the actual Center. Instead of using $(l,b)$ one can use $(l^*,b^*)$, where the center is shifted to Sgr A*, see e.g. \citet{fritz16}.

\begin{figure}
  \centering
\epsscale{1.00}
\includegraphics[trim={0cm 0cm 0cm 0cm},clip,angle=0,width=1.00\hsize]{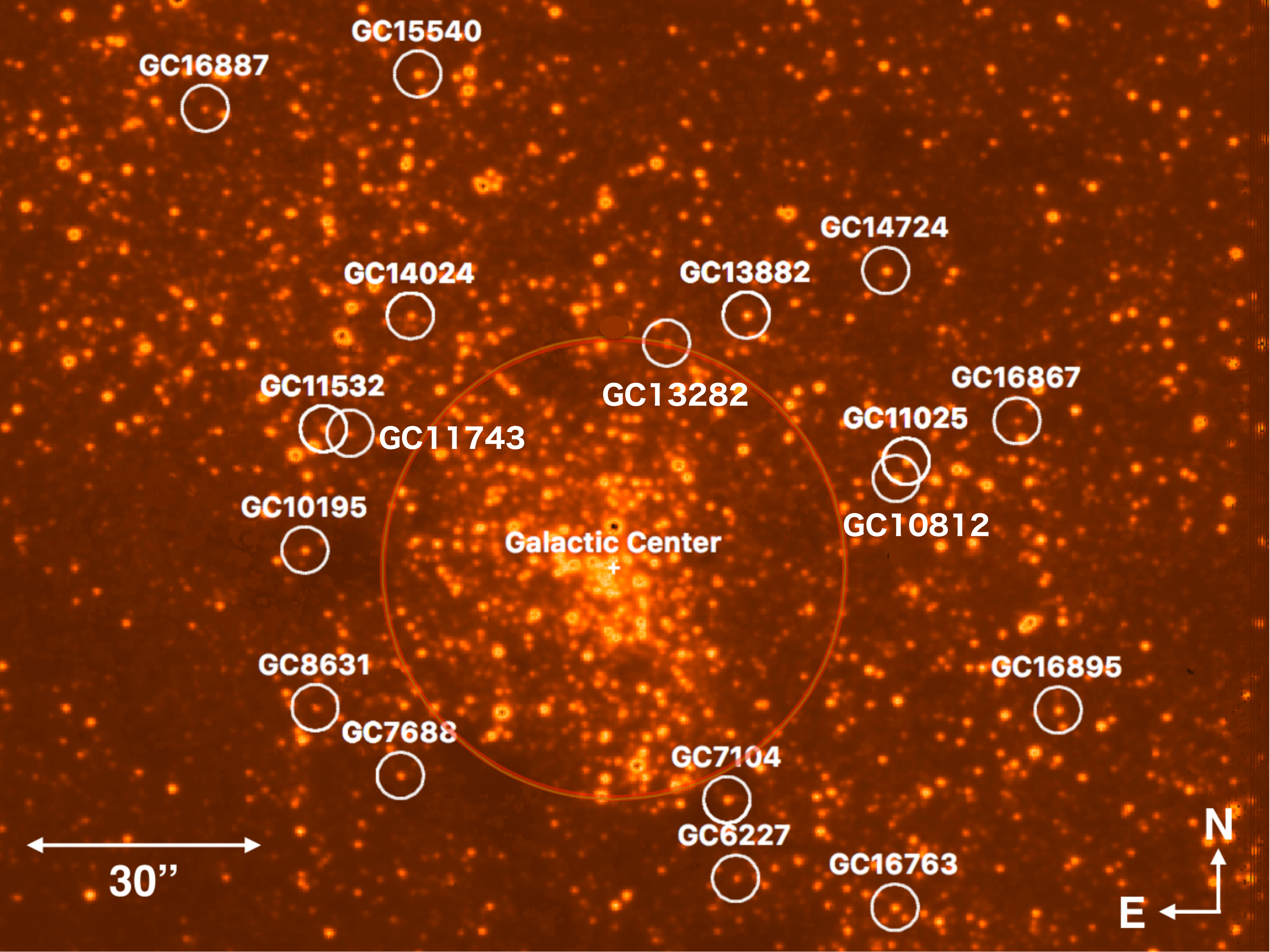} 

\caption{UKIDSS finding chart of the Galactic Center region \citep{ukidss}. The image captures the inner part of the Nuclear Star Cluster, which has a diameter of $6$'. Our target stars are marked with white circles and are summarized in Table \ref{tab:star}. The orange circle surrounding the Galactic Center denotes an angular radius of $30$\arcsec and marks the region where we have avoided stars due to problems with crowding in seeing-limited observations.  $30$\arcsec\ corresponds to a projected galacto-centric distance of $R_c=1.2$\,pc, adopting the distance to the Galactic Center of $8.3$\,kpc \citep{GCdistance,bland:16}. The white cross marks the location of the Galactic Center. \label{fig:ds9}}
\end{figure}

In Figure \ref{fig:ds9} the stars are shown on a UKIDSS finding chart \citep{ukidss}, with all of them lying within $90$\arcsec\ angular distance from the Galactic Center.  This corresponds to a projected galactocentric distance of $R_c<3.6$\,pc, adopting the distance to the Galactic Center of $8.3$\,kpc \citep{GCdistance,bland:16}.
The stellar population of the Nuclear Cluster has a half-light radius of approximately $3'$ \citep{fritz16}, exceeding by a factor of two the image shown in Figure \ref{fig:ds9}.

\begin{figure}
 \centering
\epsscale{1.00}
\includegraphics[trim={0cm 0cm 0cm -2cm},clip,angle=0,width=\hsize]{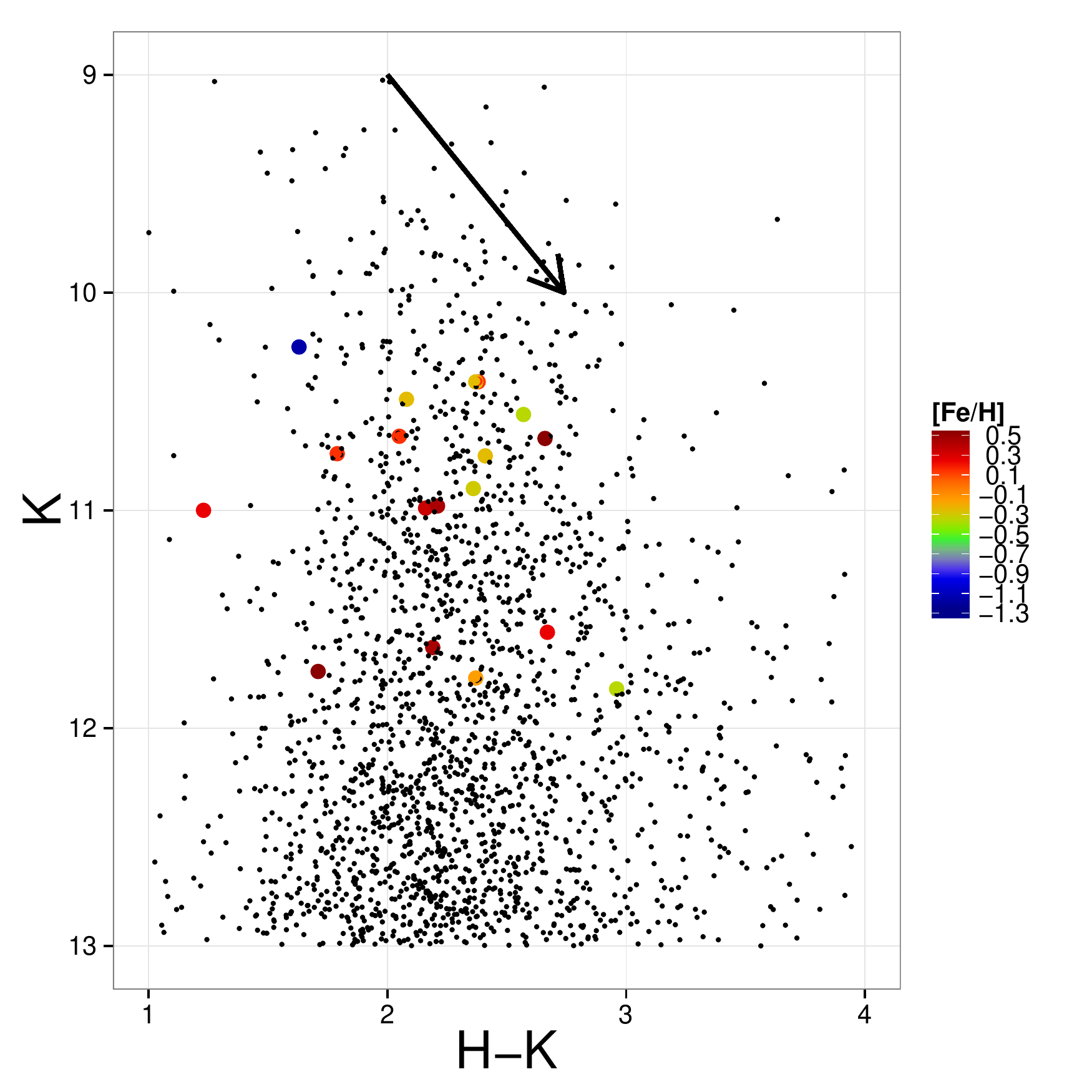} 
\caption{$K_s$ magnitude versus $H-K_S$ color for stars toward the Nuclear Star Cluster. Our target stars are marked with colored dots scaled with metallicity. The extinction vector is shown with the arrow. The most metal poor star is the same as in \citet{ryde:16}.
\label{fig:khk}}
\end{figure}

Our stars were selected from a sample of Galactic Center giants observed at low spectral resolution ($R=4000$ or $R=1500$) with the integral field spectrometer SINFONI \citep{sinfoni,sinfoni2} on the VLT\footnote{The SINFONI spectra are from the programs 077.B-0503, 183.B-0100 and 087.B-0117.}. We chose candidates with measurable CO bandheads in order to derive the stellar temperature, excluding candidates with neighbors too close for seeing-limited spectroscopy.
The orange circle in Figure \ref{fig:ds9} shows a circle of a radius of 30", enclosing a region excluded from targeting due to extreme crowding. We also used the catalogs of \citet{Blum03} and \citet{matsunaga09} to exclude some known AGB/LPV stars like Miras, and known red supergiants. The aim of this sample is to collect a 
relatively unbiased, large sample of old Galactic Center stars, which can be used in our studies of the metallicity distribution. Since we have to avoid very cool stars ($T<3300$~K) due to the difficulty of analyzing such stars, we might be slightly underestimating the mean metallicity, since these cool giants would sample the more metal-rich end. 
Our final sample should be relatively unbiased in color, covering the full $H-K_s$ range, see Figure \ref{fig:khk}. It is also encouraging that the metallicity distribution is not very sensitive to the selection function, as shown by \citet{nandakumar:17} when investigating its influence on different  spectroscopic surveys, such as APOGEE, Gaia-ESO, Lamost, and Rave.   In order to avoid an additional layer of complexity in target selection, we did not add proper motion as an additional constraint. 
Table \ref{tab:phot} gives the absolute $K$-magnitude ($M_K$\footnote{These are based on the apparent $K_s$-magnitudes, derived metallicities, and 10\,Gyr isochrones \citep{bressan:12}. The uncertainty of GC7688, which is a foreground star (see Sect. \ref{membership}), is larger due to the uncertainty in distance (1 sigma range from $4$ to $8.3$\,kpc).}, with uncertainties of $\sim 0.09$, except $\delta M_K[\mathrm{GC7688}]\sim0.8$), apparent $K_s$ magnitudes (with uncertainties of $\sim 0.05$), and $H-K_s$ colors (with uncertainties of $\sim 0.08$). NIRSPEC spectra of giants in the K band as faint as $K_s<12$ are possible to acquire with the Keck telescope, for reasonable exposure times (see Table \ref{tab:star}).

\begin{figure}
  \centering
\epsscale{1.00}
\includegraphics[trim={0cm 0cm 0cm 0cm},clip,angle=-90,width=0.9\hsize]{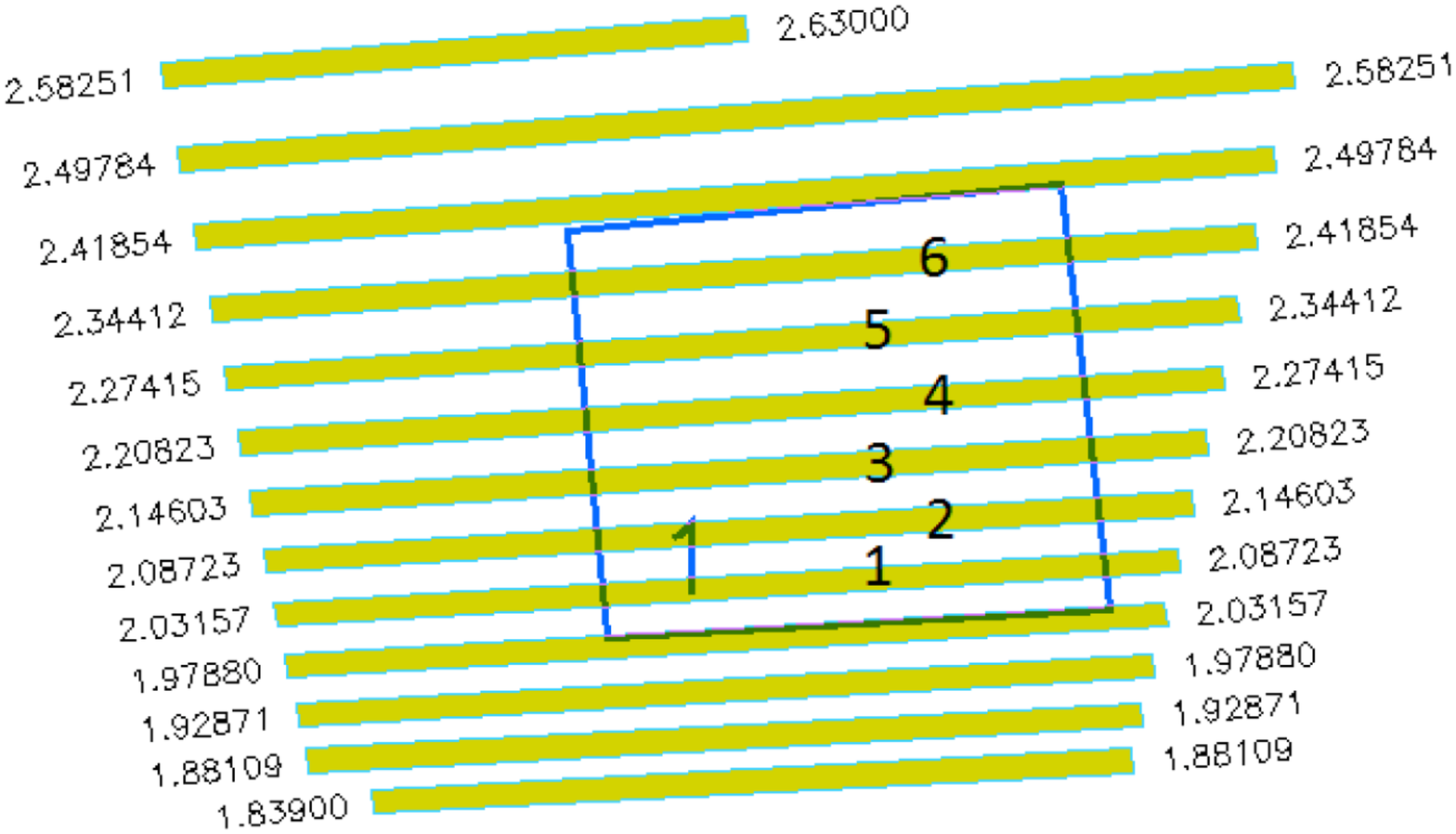} 
\caption{Outline of the spectral orders is shown by yellow bands. The detector array is marked with the blue square, which shows the spectral coverage obtained with the NIRSPEC settings used during our observations. The six spectral orders captured by the array are marked. Only orders $2-5$ were used in our analysis. \label{fig:nirspec}}
\end{figure}

Using the $0.432"\times 12"$ slit and the NIRSPEC-7 filter, a spectral resolution of $R\sim 24,000$ was achieved. The analyzed spectra cover the wavelength region from  21,100 to 24,000 \AA, distributed over 5 spectral orders with gaps in between, see Figure \ref{fig:nirspec}.  To achieve proper background and dark subtraction, we observed with an ABBA scheme with a nod throw of $6\arcsec$ along the slit. Total exposure times range $960-3000\,$s. 

The data were reduced with the NIRSPEC software {\tt redspec} \citep{nirspec_reduction} providing final 1-D wavelength-calibrated spectra. Special care was taken to ensure the tracing of the correct target star on the slit, since several stars were inevitably observed at the same time in these crowded fields. IRAF \citep{IRAF} was subsequently used to normalize the continuum, eliminate obvious cosmic ray hits, and correct for telluric lines. For the latter, we observed a few bright telluric standard stars using the same setup as the target stars, twice per night. The star finally chosen for the telluric line division was the rapidly rotating O6.5V star HIP89584, a star with a K magnitude of 7.3 and a $v(\sin i)\sim 140$\,\kms\ \citep{vsini}. The broad Brackett $n=4-7$ hydrogen line at 21655\,\AA, was smoothed out and rectified. The telluric star spectrum, observed to be used for the final night's targets, had to be filtered for a fringing pattern, using standard tools in {\tt redspec}.  

Since the telluric absorption spectrum consist of different molecular species and their scaling to the time of observations is not always the same for all species, a final check of the performance of the telluric line reduction was done by overplotting the telluric spectrum on the target. This is important in order to explain possible spurious features arising from the reduction. Also, knowledge of whether a telluric line affects the Fe lines under study is crucial. 
Special care was also taken to, if needed, eliminate the portions of the spectrum where the sky emission lines of OH were present as they cannot always be corrected. 

In order to benchmark the performance of our analysis method (see Section \ref{analysis}), we have also analyzed the KPNO high-resolution spectrum of the metal-poor giant $\alpha$ Boo \citep[Arcturus][]{aboo_atlas}, with the fundamental parameters \teff$=4286$\,K, \logg$=1.64$, \feh$=-0.57$, and $\xi_\mathrm{micro}=1.58$\,\kms\ \citep{jonsson:2017_I}.  

\begin{deluxetable*}{l c c c c c}
\tablecaption{Stellar coordinates, position, and exposure times \label{tab:star}}
\tablewidth{0pt}
\tablehead{
\colhead{Star} & \colhead{observation date} & \colhead{RA} &  \colhead{dec} &  \colhead{$(l,b)$} & exposure time  \\
   & & \colhead{[h:m:s]} & \colhead{[d:m:s]} &  \colhead{[$^\circ,^\circ$]} & [s]  
  } 
\startdata
GC10812 & 2015 Apr 27 & 17:45:37.229  & -29:00:16.62 & (359.942, -0.036) & 1200 \\
GC11473& 2015 Apr 27 & 17:45:42.642 & -29:00:10.23  &(359.953, -0.052)& 1600\\
GC13282& 2015 Apr 27 & 17:45:39.492 & -28:59:58.74  &(359.950, -0.040)&  960    \\
GC14024& 2015 Apr 27 & 17:45:42.026 & -28:59:54.97  &(359.956, -0.047)&  960     \\
GC16887& 2015 Apr 27 & 17:45:44.041 & -28:59:27.66  &(359.966, -0.050)&   2000    \\
GC6227& 2015 Apr 27 &  17:45:38.855 & -29:01:08.71  &(359.932, -0.048)&  2000     \\
GC7688& 2015 Apr 27 &  17:45:42.173 & -29:00:54.99  &(359.942, -0.057)&    960   \\
GC15540& 2015 Apr 28 & 17:45:41.931 & -28:59:23.39  &(359.963, -0.043)&   500    \\
GC10195& 2016 Apr 18 & 17:45:43.103 & -29:00:25.45  &(359.951, -0.055)&   2000  \\
GC11532& 2016 Apr 18 & 17:45:42.898 & -29:00:09.60  &(359.954, -0.052)&  2000  \\
GC13882& 2016 Apr 18 & 17:45:38.701 & -28:59:55.15  &(359.949, -0.037)&  3000  \\
GC14724& 2016 Apr 18 & 17:45:37.310 & -28:59:49.43  &(359.948, -0.032)&   2000    \\
GC16763& 2016 Apr 18 & 17:45:37.288 & -29:01:12.67  &(359.928, -0.044)&   2000    \\
GC7104& 2016 Apr 19 &  17:45:38.943 & -29:00:58.44  &(359.935, -0.047)&  2000     \\
GC11025& 2016 Apr 19 & 17:45:37.126 & -29:00:14.39  &(359.942, -0.035)&  2000    \\
GC16867& 2016 Apr 19 & 17:45:36.021 & -29:00:09.20  &(359.941, -0.031)&  2000    \\
GC16895& 2016 Apr 19 & 17:45:35.640 & -29:00:47.00  &(359.931, -0.035)&  2000    \\
GC8631  & 2016 Apr 19& 17:45:43.016 & -29:00:46.02  &(359.946, -0.058)&   3000   \\
\enddata
\end{deluxetable*}

\begin{deluxetable*}{l c c c c | c c c | c c c | c }
\tablecaption{Stellar photometry and parameters \label{tab:phot}}
\tablewidth{0pt}
\tablehead{
\colhead{}   & \colhead{} & \colhead{} & & & \multicolumn{3}{|c|}{photometry} & \multicolumn{3}{c|}{isochrones} &\\
\colhead{Star} & \colhead{$M_K$} & \colhead{$K_S$} & \colhead{$H-K_S$} &  \colhead{$T_{\mathrm{eff}}$} & \colhead{$\log g$} &  \colhead{\feh} & \colhead{$\xi_{\mathrm{mic}}$} & \colhead{$\log g$} &  \colhead{\feh} & \colhead{$\xi_{\mathrm{mic}}$}  & \colhead{$\xi_{\mathrm{mac}}$}  \\
 & & & & \colhead{[K]} & \colhead{(dex)} & \colhead{(dex)} & \colhead{[\kms]} &\colhead{(dex)}  & \colhead{(dex)}  & \colhead{[\kms]}  & \colhead{[\kms]}   } 
\startdata
GC10812 & $-6.38$ & 10.25 & 1.63 &  3800& 0.47 & $-1.18$ & 2.4 & 0.34 &  $-1.15$  & 2.5  & 14.0  \\
GC11473 & $-4.81$ & 11.74 & 1.71 &  3550& 0.97 & $+0.48$ & 2.1 & 1.13 &  $+0.64$  & 2.0  & 12.0   \\
GC13282 & $-6.24$ & 10.99 & 2.16 &  3700& 0.45 & $-0.11$ &  2.4 & 0.97 &  $-0.05$  & 2.1  & 14.0   \\
GC14024 & $-6.13$ & 10.98 & 2.21 &  3650& 0.42 & $+0.02$ & 2.4 & 1.23 &  $+0.42$  & 2.0  & 14.0   \\
GC16887 & $-5.52$ & 11.63 & 2.19 &  3500& 0.67 & $+0.24$ & 2.3 & 0.96 &  $+0.41$  & 2.1  & 13.0   \\
GC6227  & $-6.43$ & 11.82 & 2.96 &  3800& 0.75 & $-0.64$ & 2.2 & 0.95 &  $-0.38$  & 2.1  & 14.0   \\
GC7688  & $-4.34$ & 11.00 & 1.23 &  4150& 1.02 & $-0.26$ & 2.1 & 1.78 &  $-0.08$  & 1.8  & 13.5  \\
GC15540 & $-6.46$ & 10.49 & 2.08 &  3600& 0.28 & $-0.19$ &  2.5 & 0.72 &  $-0.16$  & 2.2  & 14.0   \\
GC10195 & $-7.15$ & 10.67 & 2.66 &  3500& 0.19 & $-0.03$ & 2.6 & 0.68 &  $+0.04$  & 2.3  & 13.0 \\
GC11532 & $-6.59$ & 10.90 & 2.36 &  3500& 0.29 & $-0.47$ & 2.5 & 0.35 &  $-0.56$  & 2.5  & 14.0  \\
GC13882 & $-7.02$ & 10.41 & 2.37 &  3300& 0.15 & $-0.31$ & 2.6 & 0.15 &  $-0.31$  & 2.6  & 15.5 \\
GC14724 & $-7.18$ & 10.56 & 2.57 &  3600& 0.16 & $-0.44$ & 2.6 & 0.55 &  $-0.49$  & 2.3  & 14.0   \\
GC16763 & $-6.35$ & 10.66 & 2.05 &  3600& 0.29 & $-0.23$ & 2.5 & 0.65 &  $-0.29$  & 2.3  & 14.0  \\
GC7104  & $-5.96$  & 10.74 & 1.79 &  3750& 0.53 & $-0.31$ & 2.4 & 0.92 &  $-0.28$  & 2.1  & 14.0 \\
GC11025 & $-6.88$ & 10.41 & 2.38 &  3400& 0.21 & $+0.34$ & 2.6 & 0.69 &  $+0.27$  & 2.3  & 14.3 \\
GC16867 & $-5.73$ & 11.77 & 2.37 &  3650& 0.66 & $-0.33$ & 2.3 & 0.74 &  $-0.29$  & 2.2  & 13.0 \\
GC16895 & $-6.69$ & 10.75 & 2.41 &  3450& 0.23 & $-0.16$ & 2.6 & 0.51 &  $-0.10$  & 2.4  & 14.0    \\
GC8631  & $-6.18$ & 11.56 & 2.67 &  3900& 0.61 & $+0.18$ & 2.3 & 1.63 &  $+0.36$  & 1.9  & 13.5  \\
\enddata
\end{deluxetable*}

\begin{deluxetable*}{l c c c c}
\tablecaption{Stellar kinematics \label{tab:kin}}
\tablewidth{0pt}
\tablehead{ \colhead{Star} & \colhead{$v^{\mathrm{LSR}}_{\mathrm{rad}}$}    & \colhead{$\mu_l$}  & \colhead{$\mu_b$} & r \\
\colhead{}  & \colhead{[\kms]}  & \colhead{\,mas\,yr$^{-1}$} & \colhead{\,mas\,yr$^{-1}$} & \colhead{[$\arcsec$]} }
\startdata
GC10812 & -44  	& 2.31$\pm$0.27 & -3.12$\pm$0.27& 39 	 \\
GC11473 &  206 	& 				&     			& 39    \\
GC13282 &  128 	& -2.85$\pm$1.95& -2.12$\pm$1.95& 30   \\
GC14024 &  -16 	& 2.08$\pm$0.21 & 1.90$\pm$0.21 & 42    \\
GC16887 &  60  	& 				& 				& 80 		 \\
GC6227  & -102 	& -2.83$\pm$0.58& 1.02$\pm$0.58 & 43 	\\
GC7688  &  -22 	& 0.40$\pm$0.71 & 0.85$\pm$0.71 & 38 		\\
GC15540 & -54 	& -2.07$\pm$0.09& 0.27$\pm$0.09 & 69 	 \\
GC10195 & $63$ 	& $0.44\pm0.70$ & $0.75\pm0.70$ & $40$   \\
GC11532 &  -28 	& 2.08$\pm$0.74 & 1.18$\pm$0.74 & 42    \\
GC13882 &  195 	& -1.66$\pm$0.81& -1.78$\pm$0.81& 37   \\
GC14724 &  -3 	& -3.46$\pm$0.82& 1.13$\pm$0.82 & 53 	 \\
GC16763 &  -44 	& 				&  				& 57 	  \\
GC7104  &  -58 	& 1.76$\pm$0.23 &0.19$\pm$0.23  & 34 		\\
GC11025 & 67  	& 				&      			& 41   \\
GC16867 &  38 	& 				& 				& 56 	\\
GC16895 &  -95  & 				&  				& 61 		\\
GC8631  &  -20 	& 				&   			& 43
\enddata
\end{deluxetable*}

\begin{figure}
  \centering
\epsscale{1.00}
\includegraphics[trim={0cm 0cm 0cm 0cm},clip,angle=0,width=1.00\hsize]{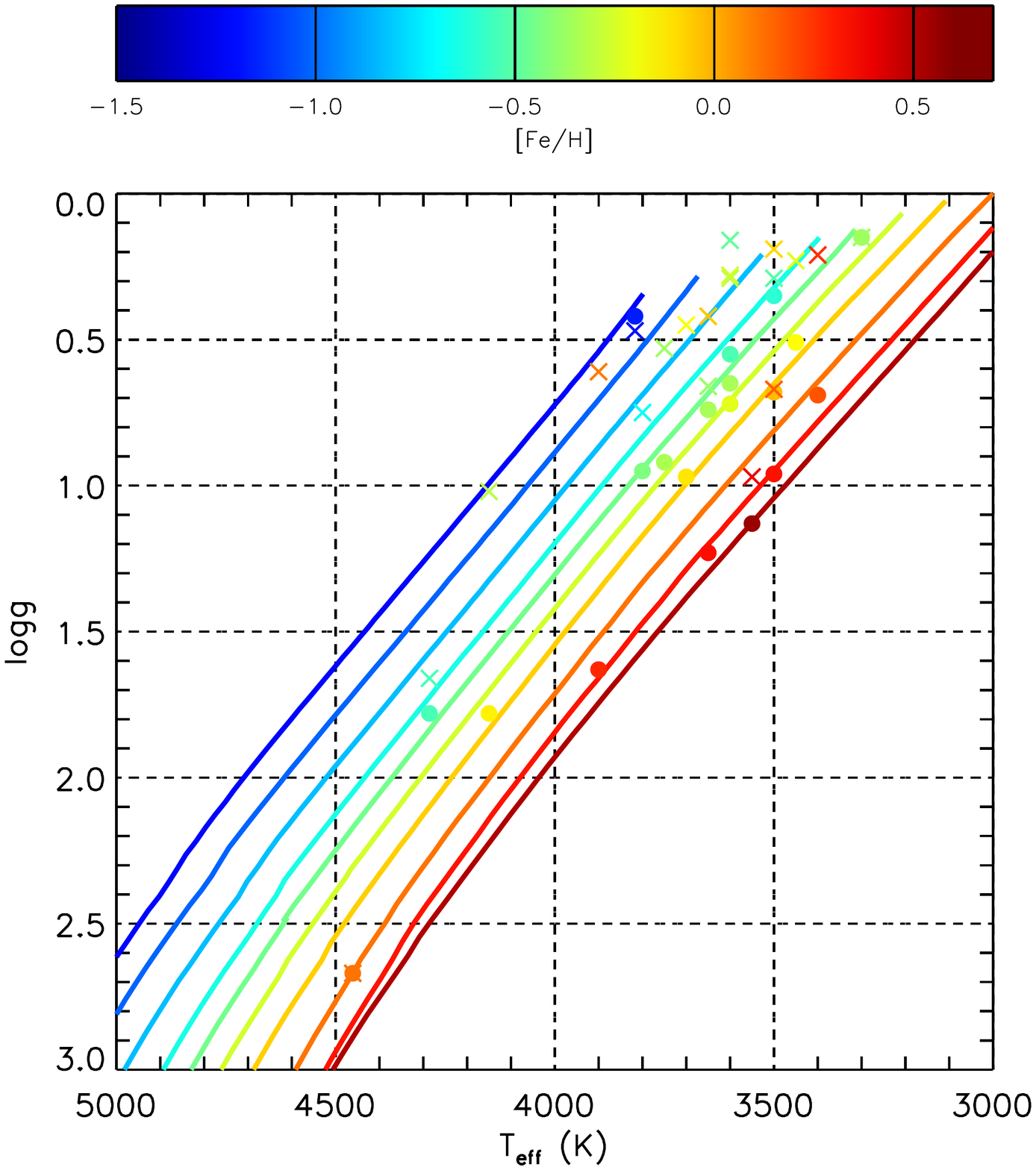} 
\caption{HR diagram Yonsei-Yale isochrones \citep{yy} with an appropriate [$\alpha$/Fe] enhancement for metal-poor stars. The metallicities of the isochrones are from [Fe/H]=$-1.2$ to $0.6$, in steps of 0.2 dex. The \logg\ of the stars are determined given the \teff\ determined from low-resolution spectra and the metallicity from the high-resolution spectra, assuming an old age of 3-10 Gyrs. The two warmest giants  are the standard stars $\alpha$ Boo (Arcturus) and $\mu$ Leo.
\label{fig:yy}}
\end{figure}

\begin{figure}
  \centering
\epsscale{1.00}
\includegraphics[trim={0cm 0cm 0cm 0cm},clip,angle=0,width=1.00\hsize]{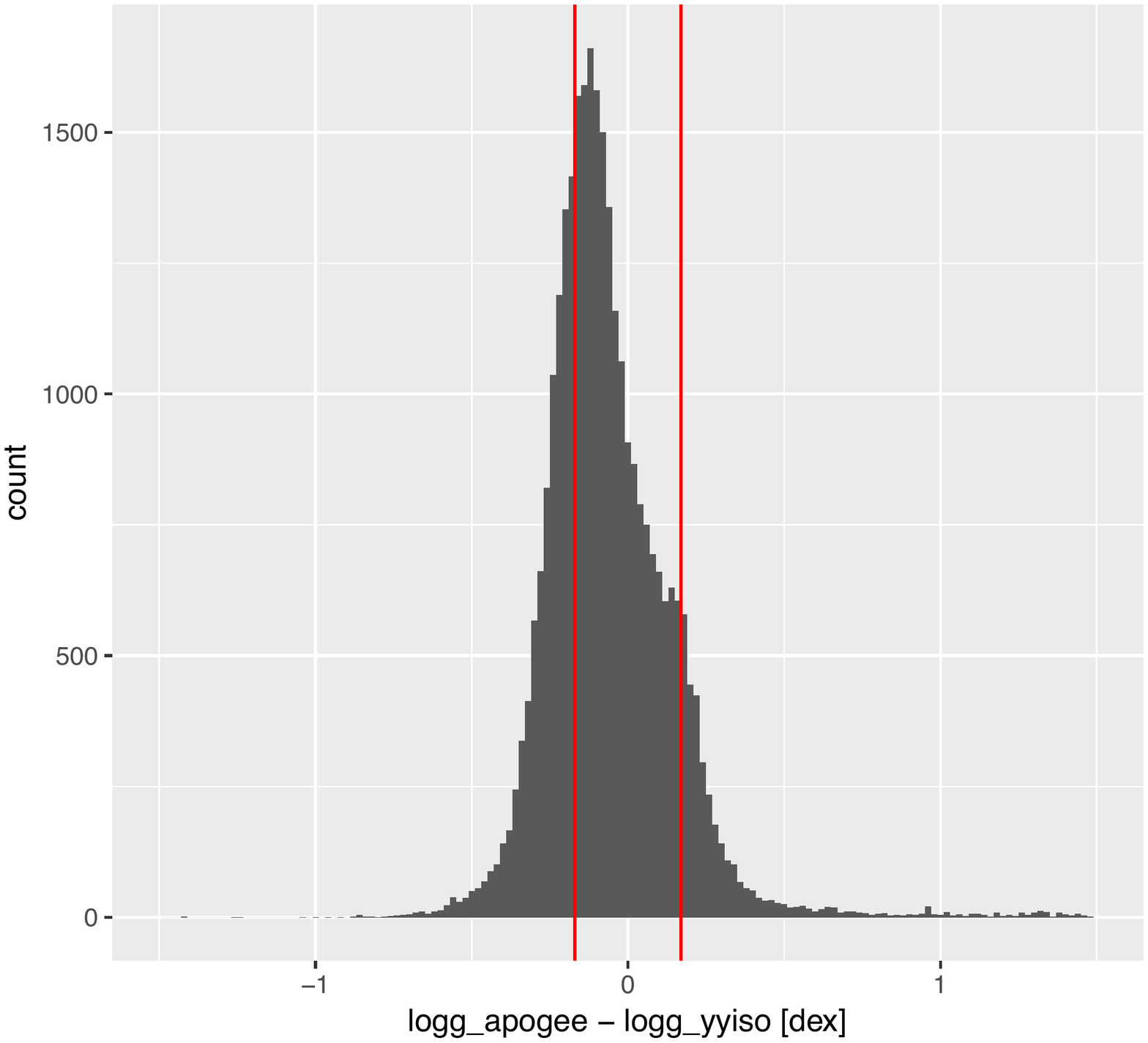} 
\caption{Histogram of the difference between \logg\ determined by  the APOGEE consortium for 30\,000 giant stars in the APOGEE DR13 sample  and \logg\ determined by our isochrone-fitting method. The vertical lines represent the $\pm2\sigma$ interval using the uncertainties reported by APOGEE, with $\sigma = 0.08$.}
\label{fig:loggdiff}
\end{figure}

\begin{figure}
 \centering
\epsscale{1.00}
\includegraphics[trim={0cm 0cm 0cm 0cm},clip,angle=0,width=1.0\hsize]{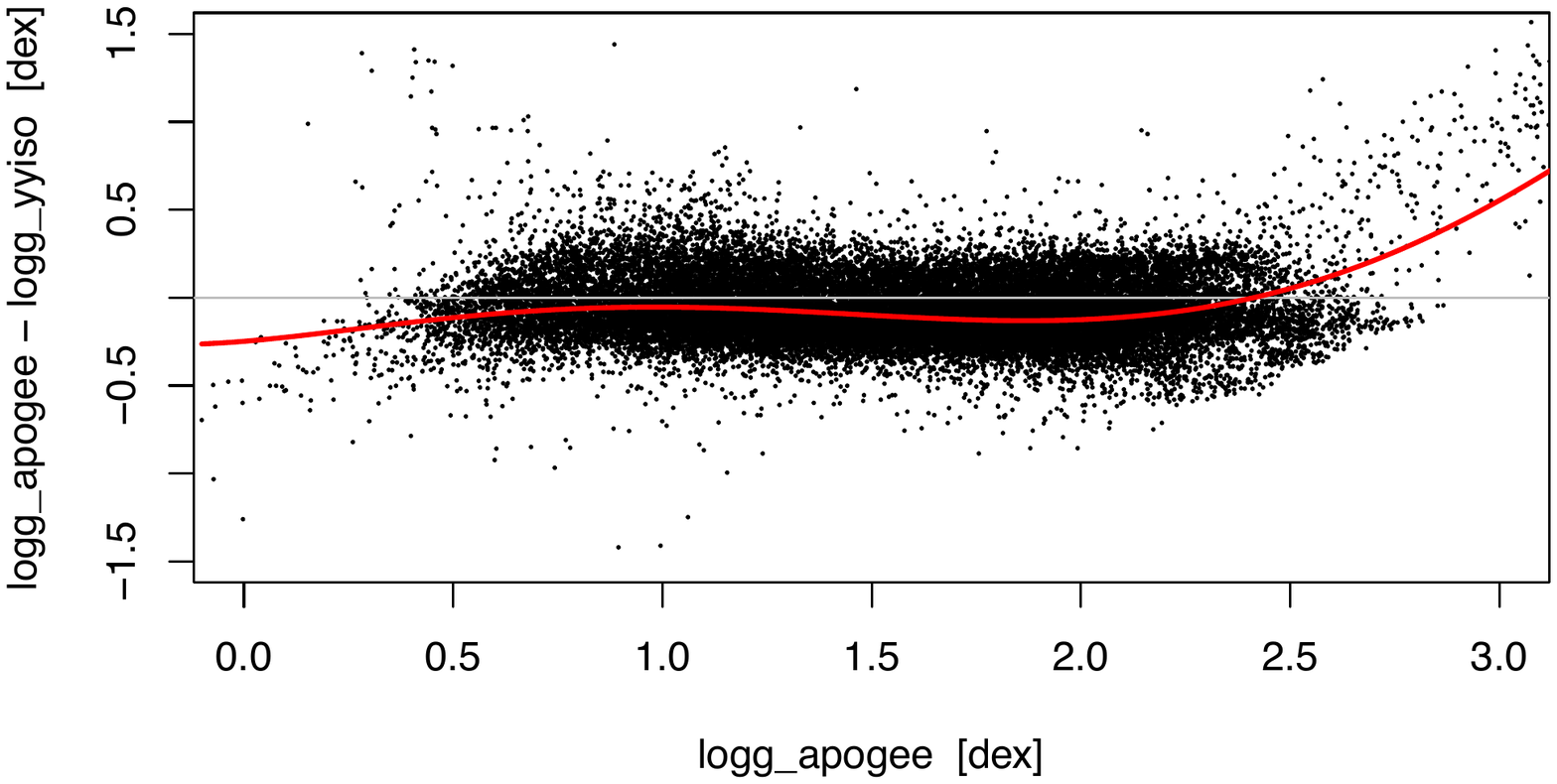} 
\caption{Difference between the determination of \logg\ by  the APOGEE consortium for 30\,000 giants and that by our isochrone-fitting method is plotted versus APOGEE's \logg. The red line is the 0.50 quantile for the data. \label{fig:diffvslogg}}
\end{figure}

\section{Analysis}
\label{analysis}

\begin{figure*}
 \centering
\epsscale{1.00}
\includegraphics[trim={0cm 2.5cm 0cm 0cm},clip,angle=90,width=1.0\hsize]{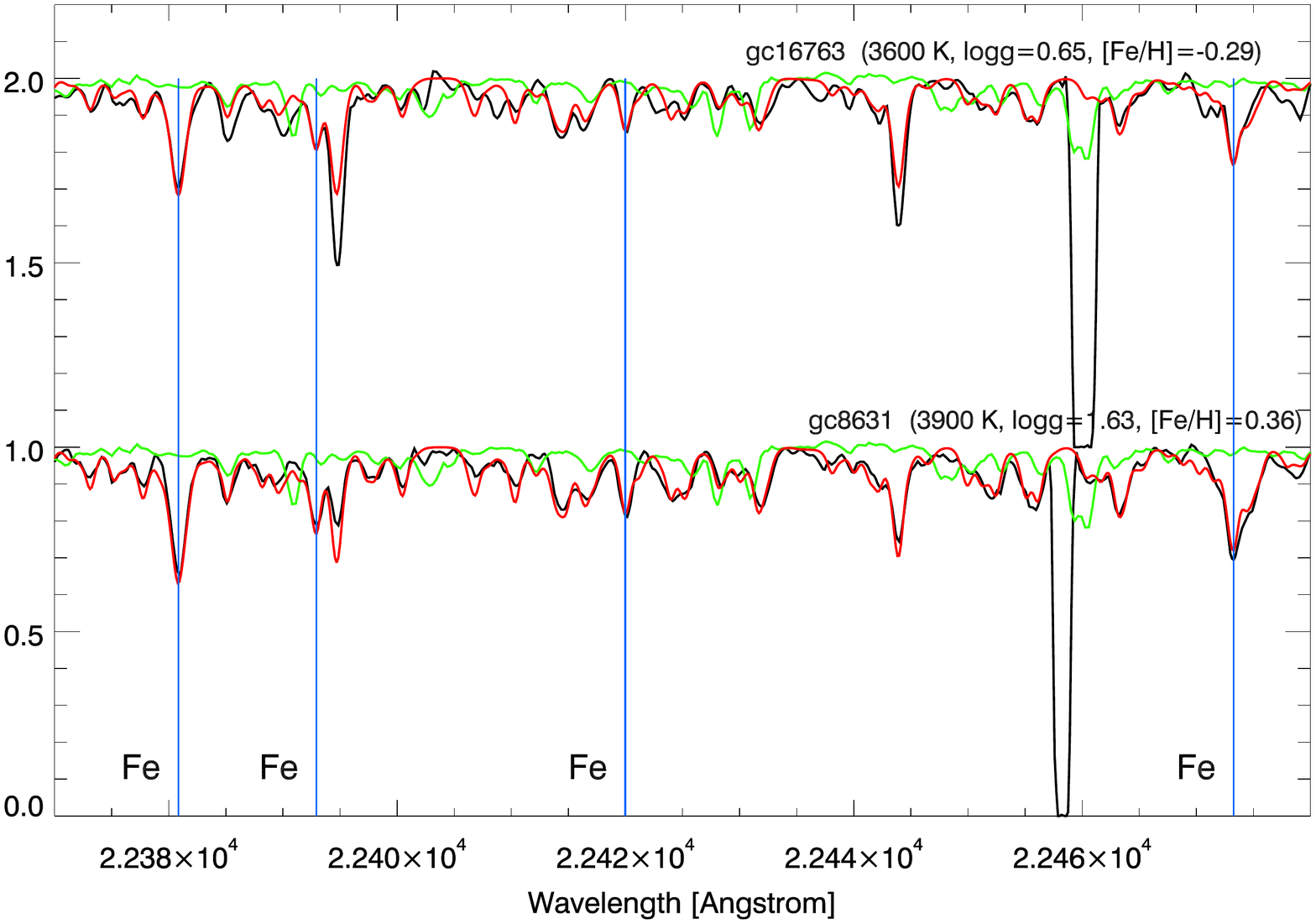} 
\caption{Two normalized, high-resolution spectra of giants in the Nuclear Star Cluster, one metal-poor and one metal-rich. The metal-poor spectrum is shifted up for clarity.
Four Fe lines are marked with the blue lines. Synthetic spectra are shown in red and the green spectrum shows where the telluric lines hits the spectrum. Apart from the strong Sc line red-ward of the second Fe line and a Ti line at $22444$\,\AA\, the other spectral features are mainly due CN lines.  \label{fig:twospectra}}
\end{figure*}

\begin{figure*}
 \centering
\epsscale{1.00}
\includegraphics[trim={2cm 1.5cm 1cm 2cm},clip,angle=0,width=0.67\hsize]{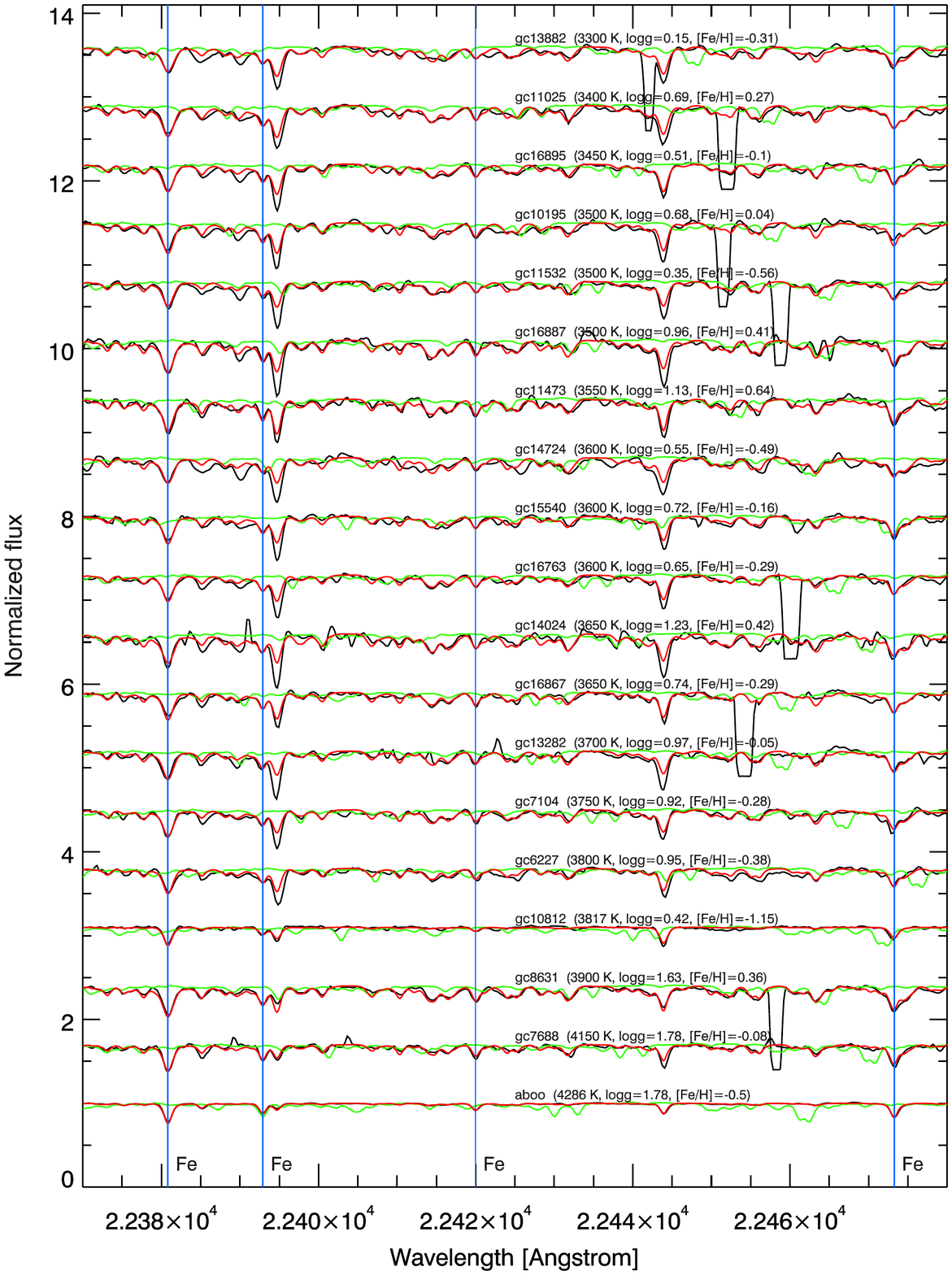} 
\caption{17 high-resolution spectra of giants in the Nuclear Star Cluster as well as that of GC7688, and the reference stars Arcturus ($\alpha$ Boo). Names and stellar parameters are indicated above every spectrum. Four Fe lines are marked with the blue lines. Synthetic spectra are shown in red and the green spectrum shows where the telluric lines hits the spectrum. The missing parts of a few of the spectra are wavelength regions where telluric emission lines of OH lie. Sometimes these are not possible to divide out. Apart from the strong Sc line red-ward of the second Fe line and a Ti line at $22444$\,\AA\, the other spectral features are mainly due CN lines.  \label{fig:spectra}}
\end{figure*}

 In this first paper of the series, we investigate the Fe abundances of stars in the Nuclear Star Cluster, in order to measure its metallicity distribution. In forthcoming papers, we will also investigate other elements; most significantly we will investigate the diagnostically important $\alpha$-element abundance trends with metallicity. Several $\alpha$ elements are represented in the wavelength region we have observed (namely Mg, Si, Ca, and Ti). However, with respect to the number of lines, line strengths, and available non-LTE corrections, our ongoing work indicates that the Si abundance is the best proxy for the general $\alpha$-element trends. The other elements might be more challenging. Furthermore, the very strong Sc lines detected in our stars are of great interest to investigate as to whether or not their strengths actually reflect a higher-than-normal Sc-abundance. These issues will be addressed in forthcoming papers in the series.

 In all our investigations, we have analyzed (and will analyze) our red-giant spectra with tailored synthetic spectra, based on spherical model atmospheres. Due to the extreme extinction and its patchy nature toward the Galactic Center,  the effective temperatures and surface gravities cannot be determined photometrically without unacceptably high uncertainty. Therefore, we have to resort to other methods, described in the following subsections.  With the stellar parameters determined, we determine  the metallicity from 7  individual Fe lines in the K band, by minimizing the $\chi^2$ between the observed and synthetic spectra.

\subsection{Stellar Dynamics}

The angular distance from Sgr~* and the LSR radial velocities ($v_\mathrm{rad}^\mathrm{LSR}$) of all our stars are listed in Table~\ref{tab:kin}. We also provide proper motions ($\mu_l$ and $\mu_b$) for those stars for which we were able to retrieve proper motions (see discussion in Sect. \ref{membership}).

\subsection{Spectral Synthesis}

The main input for tailoring the synthetic spectra are the stellar parameters and an accurate line list, in this case of Fe lines. The accurate determination of the stellar parameters of the target stars is of large importance for the final result and its uncertainties.  The four fundamental parameters (or stellar parameters) are the effective temperature (\teff), surface gravity (\logg),  metallicity ([Fe/H]), and microturbulence ($\xi_\mathrm{mic}$).  Occasionally, $[\alpha/\rm{Fe}]$ (especially [Mg/Fe] is also considered a fundamental parameter, due to its influence on the continuous opacity. In the following subsections we will discuss these parameters and the line list we use, before discussing the abundance determination itself.

\subsubsection{Effective Temperatures}

The effective temperatures, $T_\mathrm{eff}$, of the target stars, are determined by integrating the strength of the $2.3\,$\mic\ ($v=2-0$) CO band in the integral field  SINFONI spectra (at spectral resolution of $R=4000$ and $R=1500$ ) and using the relation between the CO-band strength and the \teff\ given in \citet{schultheis:16}.  These authors have shown that this relation works very well in the temperature and metallicity ranges we are interested in. It is valid between $3200$\,K and $4500$\,K and  $-1.2<$\feh$<0.5$, with a  typical dispersion of about $150$\,K. Indeed, we find no correlation between \teff\ and \feh. Our derived effective temperatures are given in the fifth column of Table \ref{tab:phot}.


\subsubsection{Surface Gravities\label{logg}}


The surface gravity, \logg, is important to set the pressure structure of the model atmospheres. Also, in the calculation of the synthetic spectra, the continuous opacity as well as the ionization states of atoms, are sensitive to it. There are several methods that can be applied to determine  the surface gravity for giants. Unfortunately, these usually require high-resolution optical spectra and employ the ionization balance between Fe{\sc i} and Fe{\sc ii} lines or the strengths of pressure-broadened wings of strong lines; these are not available for the K band.  

Instead, we initially constrain the surface gravities photometrically. Since we assume that our target stars are members of the Nuclear Star Cluster (see Sect. \ref{membership}), the distance is in principle very well determined. The distances can otherwise be a large source of uncertainty for photometrically determined \logg\ \citep[see for example][]{schultheis:17}. On the other hand, the interstellar  extinction  toward the Galactic-Center region is both high and variable \citep{schodel2015} which plagues the photometric \logg\ determination, overwhelming the other uncertainties in the stellar masses, bolometric corrections, and effective temperatures.  We determined the extinction, star by star by using the $H-K$ color excess and assuming an intrinsic $(H-K)_{0} = 0.18$. 
The color excess $E(H-K)$ was converted to $A_{K}$ using the extinction law of \citet{fritz:11} with $A_{K} = 1.351 \times E(H-K)$.   However, as shown in
 \citet{schultheis:14}, 
 this method overestimates the extinction for high $A_{K}$ values and also show some dependence on stellar parameters such as \teff, \logg, and [Fe/H]. A rough estimate of a general uncertainty
 in the photometric \logg\ is $\sim 0.3$ dex  taking into account
 the uncertainties in the extinction determination as well as in the extinction law.  The uncertainties could, however, easily  exceed that due to the large uncertainties in the extinction, $A_{K}$. The gravities determined photometrically are given in the sixth column in Table \ref{tab:phot}. 

These photometric gravities and the spectroscopically determined effective temperatures of our target stars are plotted in the HR-diagram in Figure \ref{fig:yy} as crosses, color-coded with the metallicities we derive for these stellar parameters (see Sect. \ref{feh} for details). Since calculated isochrones tell us that not all combinations of \teff, \logg, and \feh\ are possible, we have also plotted Yonsei-Yale 10 Gyr isochrones \citep{yy} appropriate for bulge stars in the Figure. This will show where our stars with given \teff\ are expected to lie, for given metallicities.  Our photometrically determined  \logg\ do not obviously fall on the corresponding isochrones. Some stars are off by as much as a factor of 10 which corresponds to overestimating the extinction by $A_{K} \sim 0.8$\,mag, an acceptable error in this region.
 Since the uncertainties are expected to be large and spurious, we have also resorted to another method for determining the surface gravities of our target stars, using the information buried in the isochrones. 

Assuming that our Galactic Center stars are all older than at least $\sim 5$\,Gyr \citep{Blum03,clarkson,Pfuhl_11} and thanks to the fact that isochrones for stars of ages $5-13$\,Gyrs are quite insensitive to the age on the giant branch, we can constrain 
the surface gravities, given a derived \teff\ from our low-resolution spectra and \feh\ from our high-resolution spectra.  
This is done in an iterative manner, in the sense that a \logg\ is assumed when determining a first guess of the metallicity. The star is then plotted in an HR-diagram and if the star falls far away from the isochrone with the relevant metallicity, the \logg\ is changed accordingly, and a new iteration is made. We have interpolated in a grid of Yonsei-Yale isochrones for giants of 10 Gyrs \citep{yy}. The  isochrones are $\alpha$ enhanced, such that [$\alpha$/Fe]=0.3 for \feh $<-0.5$, and for higher metallicities the [$\alpha$/Fe] linearly decreases with metallicity up to solar metallicity, above which it is zero.   For a given set of (\teff,\feh) we can then get the corresponding \logg\ by interpolating in the grid.  These isochrones are shown in Figure \ref{fig:yy}, where we also have color-coded them with their metallicities.

The \logg\ determination is quite sensitive to the location in the HR diagram. If a star has a determined metallicity which is, for example, too low compared to its location on the  isochrone, the \logg\ has to be lowered for a given temperature. Decreasing the surface gravity will tend to decrease the pressures, including the electron pressure in the atmospheres, which leads, in general, to stronger lines, since the continuous opacity decreases. The metallicity determined from the observed spectra for a lower \logg\ will therefore decrease even more. After a few iterations a consistent set of \teff, \logg, and \feh\ (given an age of 5-13 Gyrs) can be found.  

We have tested this  purely spectroscopic method on 30\,000 giant stars from the DR13 APOGEE sample \citep{apogee_ref}. Using the effective temperatures and metallicities derived by APOGEE, we determine the surface gravities through our isochrone-fitting method  (assuming 10 Gyr isochrones) and compare these to those derived by APOGEE.  No independently derived distances  are required to apply this method. We find a median difference of $-0.1\pm0.1$ dex between the \logg\ determined by APOGEE and the \logg\ predicted by our isochrone fitting method, see Figure~\ref{fig:loggdiff}. APOGEE reports a general uncertainty of $0.08$ for all of their \logg\ determinations, as plotted in Figure~\ref{fig:loggdiff}, consequently the derived differences cannot be significant. 
 In the histogram, there is a shoulder or maybe a bi-modality. This might, however, be 
expected since among the APOGEE stars in the solar neighborhood there are also young stars, whereas our method assumes an old age. 
Allowing for younger isochrones, our isochrone-determined $\log g$ would decrease by as much as 0.15 dex. All though we do not have 
any information on the ages of the APOGEE stars, we might  assume that the large peak in the bi-modality in Figure~\ref{fig:loggdiff}
is the young, thin disk.  If correct, using young isochrones would move that peak to the right, resulting in a more uniform distribution. 
In Figure \ref{fig:diffvslogg} we plot this difference versus \logg\ to show that there is no trend with \logg.

In \citet{jonsson:2017_I} carefully and spectroscopically determined stellar parameters of K giants are compared to the \citet{bressan:12} isochrones for different ages. The overall agreement is excellent, which again is very reassuring for our method. Comparing the Yonsei-Yale isochrones \citep{yy} to the ones of \citet{bressan:12}, we find an agreement of better than 0.1 dex for the full metallicity range for our type of giants.

The final surface gravities derived from isochrones are given in the ninth column of Table \ref{tab:phot}. Our stars are also plotted in an HR diagram in Figure \ref{fig:yy} as large dots, color-coded with their metallicities derived in Sect. \ref{feh}.

\subsubsection{Microturbulence}

The microturbulence, $\xi_\mathrm{micro}$, is introduced in stellar atmosphere modeling and spectral synthesis in order to capture non-thermal motions in the stellar atmosphere, which occur on scales smaller than a mean-free path of the photons and therefore affects the line formation process and the radiative transfer, analogously to thermal motions \citep[see e.g.][]{gray:2005}. Only the equivalent widths of strong lines are affected by the microturbulence, which is why it can observationally be set by the requirement that the abundances determined from, for example, Fe lines of a large range of line strengths, produce no trend. Since there are not enough Fe line of different strengths in our spectra, we have used a relation with \logg\ based on $\xi_\mathrm{micro}$ values derived from analysing spectra of five red giant stars ($0.5<\log g<2.5$) in this way by \citet{smith:13}, in which we interpolate. The lines we use for the metallicity determination are affected by the microturbulence.
Our estimated microturbulences are given in the eighth and eleventh columns of Table \ref{tab:phot}, for the photometrically determined \logg\ and for \logg\ fitted from isochrones, respectively.

\subsubsection{$\alpha$-element Over-abundance}

Since the line strengths will depend on the number density of free electrons in the stellar atmosphere, especially the Mg abundance might be important. We have assumed a general  $\alpha$-element enhancement typical for bulge giants such that for metallicities lower than \feh $=-0.5$ the over abundance of the $\alpha$ elements is [$\alpha$/Fe]=0.3. For higher metallicities up to solar, the [$\alpha$/Fe] linearly decreases with metallicity. Above solar it is assumed to be zero.  

\subsubsection{Line List}

\begin{deluxetable}{c c c }
\tablecaption{Fe line-list \label{tab:linelist}}
\tablewidth{0pt}
\tablehead{ \colhead{Wavelength in air, $\lambda_\mathrm{air}$} & \colhead{$E_\mathrm{exc}$} & \colhead{$\log gf$}  \\
\colhead{[\AA]}  & \colhead{[eV]}  & \colhead{(dex)} }
\startdata
21779.651 & 3.640 & -4.30 \\
21894.983 & 6.145 & -0.14 \\
22380.835 & 5.033 & -0.41 \\
22392.915 & 5.100 & -1.21 \\
22419.976 & 6.218 & -0.23 \\
22473.263 & 6.119 &  +0.48 \\
22619.873 & 4.991 & -0.36
\enddata
\end{deluxetable}

The Fe lines  used are taken from a carefully compiled and tested line list for use in the K band (Thorsbro et al. 2018, in prep.).  
This  atomic line list is based on an extraction from the VALD3 database \citep{vald,vald2,vald3,vald4,vald5}. Wavelengths and line-strengths (astrophysical $\log gf$-values) are updated using the solar center intensity atlas \citep{solar_IR_atlas}. We have also made use of recent laboratory measurements of atomic line strengths of  Mg, and Sc  \citep{pehlivan:mg,pehlivan:sc}, as well as of Si (Pehlivan et al., 2018, in prep.) in the K band.  Of approximately 700
identified, interesting spectral lines for cool stars, about 570 lines have been assigned new values. A subset of Fe lines from this line list, used in the present investigation, is given in Table \ref{tab:linelist}.

We also observed ten benchmark stars of similar stellar parameters in the same fashion as our target stars,  in order to test the performance of this line list.  We find an excellent agreement, to within $0.05$\,dex, between the abundances we determined with this list and the reference values for these stars.
In the abundance analysis, we also include molecular line-lists of CN \citep{sneden:14}.  

\subsection{Abundance Determination\label{feh}}

In the spectral synthesis, the radiative transfer and line formation is calculated for a given model atmosphere defined by the fundamental stellar parameters. We have chosen to use the code {\it Spectroscopy Made Easy, SME} \citep{sme,sme_code} for this. {\it SME} uses a grid of model atmospheres in which the code interpolates for a given set of fundamental parameters of the analyzed star. We use are 1-dimensional MARCS models, which are hydrostatic model photospheres in spherical geometry, computed assuming LTE, chemical equilibrium, homogeneity, and conservation of the total flux (radiative plus convective, the convective flux being computed using the mixing-length recipe) \citep{marcs:08}.
The program then iteratively synthesizes spectra, calculated in spherical symmetry, for the searched abundances, under a scheme to minimize the $\chi^2$ when comparing with the observed spectra.  The spectral lines, which are used for the abundance analysis, are marked with masks in the pre-normalized observed spectra. 

As in all spectral analyses, it is the contrast between the continuum and line depth that provides the abundance. Therefore, the continuum in the observed spectrum has to be well defined. The observed continuum levels depend on the signal-to-noise ratio of the spectra and possible residuals from the telluric line reduction. Therefore, special care needs to be given to scrutinize and to define local continua around the spectral lines used for the abundance determination.  Furthermore, the continuous opacity in the  spectral synthesis has to be correct. In the atmosphere of a cool star, the continuous opacity is due to H$^{-}$ free-free opacity, which is affected by the electron density.  The abundances of the species that are the major electron donors in the continuum forming regions need to be determined as well. For a typical star in our sample, the major electron donors are magnesium (1/2 of all electrons), iron (1/3 of electrons),  and Si (1/10); this is why the $\alpha$-element abundances is important  as an input parameter. 


Due to the way SME works, by minimizing the $\chi^2$ when comparing observed spectra with synthesized spectra at every sampling point, the line profiles need to be accurately characterized and the broadening of the line must be well determined. At the resolution of the instrument, the broadening is mainly due to the spectrometer's resolution, which we can determine from narrow, unsaturated telluric lines to be $R\sim 24,000$. The intrinsic broadening of the stellar lines is, however, also important. Of the stellar-line broadening mechanisms, the macroturbulence ($\xi^\mathrm{stellar}_\mathrm{macro}$), due to large-scale motions  in the stellar atmosphere which reshape the lines, is unknown but can be determined by fitting stellar lines. The total Gaussian broadening needed to broaden the synthetic lines ($\xi^\mathrm{total}_\mathrm{macro}$), which includes the instrumental profile, are estimated from well-formed, medium-weak lines. The final values are given in Table \ref{tab:phot} as $\xi_\mathrm{macro}$ for short.


Molecular lines are ubiquitous in spectra of cool stars of the type we are investigating.  CN is the most  dominant molecule apart from the CO bandhead region. We have fitted these lines in order to fit the over-all spectrum but also to take care of possible minor blends in the Fe lines.

Assumptions that {\it Local Thermodynamic Equilibrium, LTE} is valid are made when synthesizing the spectra of Fe lines.  \citet{nlte_fe} show that this is a good approximation for cool, low-gravity stars, where the photoionization, which is the main cause for departure of LTE, is unimportant for the statistical equilibrium of neutral iron. 

A remaining question is whether 1-D model atmospheres are still appropriate.
\citet{cerni:17} show for their red giants, which are slightly warmer than ours, that abundance corrections when using a full three-dimensional  hydrodynamical line modeling are small compared to using a traditional 1-D approach, which means that the influence of convection should be small.

In Table \ref{tab:phot} we  present the metallicities derived for our stars  for both sets of  \logg.

\subsection{Uncertainties in the  Derived Abundances}

The uncertainties in the derived metallicities are allocated in part between the fitting procedure of the observed to the synthetic spectrum,  and the uncertainties in the determined stellar parameters. 

\begin{figure}
 \centering
\epsscale{1.00}
\includegraphics[trim={0cm 0cm 0cm -1cm},clip,angle=90,width=1.1\hsize]{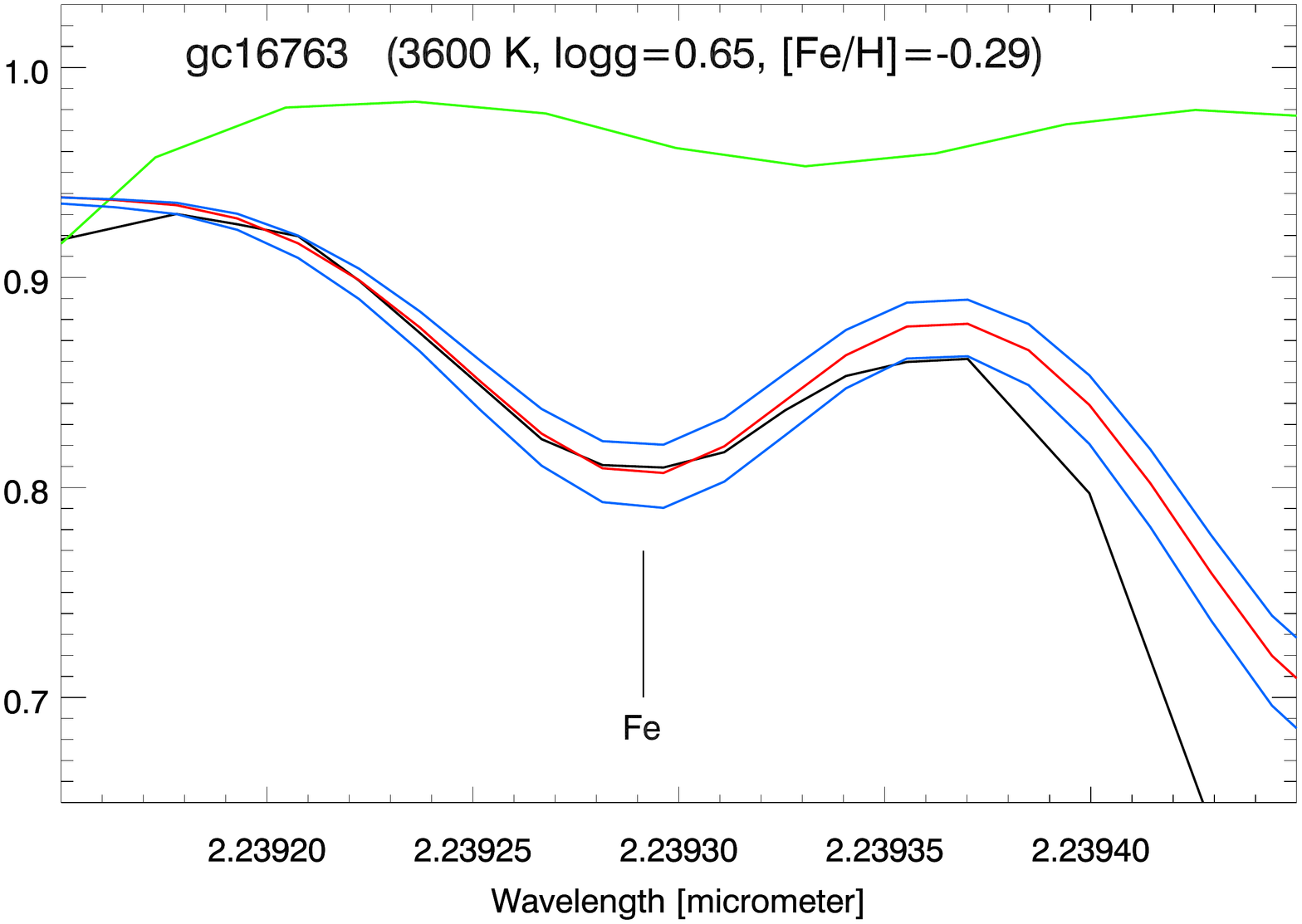} 
\caption{A zoom-in of the GC16763 spectrum,
also shown in Figures \ref{fig:twospectra} and \ref{fig:spectra}. 
 The observed spectrum of one of the iron lines used in the metallicity determination is shown in black and the best synthetic spectrum in red, for the stellar parameters given above the spectrum. Overlaid are two synthetic spectra in blue, one with $\Delta$[Fe/H]$=+0.2$ and one with  $\Delta$[Fe/H]$=-0.2$. This shows the sensitivity of the spectral lines to a change of this magnitude. The green spectrum is the telluric spectrum.
\label{fig:error}}
\end{figure}

Noise in the spectra, telluric residuals, and level of molecular contamination in the continuum make the continuum levels uncertain. Also, at $R=24,000$ the sensitivity of the lines to a change in abundance (see Figure \ref{fig:error}\footnote{In Figure \ref{fig:error} a change in metallicity by $\pm0.2$\,dex is shown and is clearly detectable.}) is naturally less than at higher resolution. We estimate these systematic uncertainties to be of the order of $\pm 0.15$\,dex in the [Fe/H] abundance.

We present the uncertainties in the metallicity determination caused by a typical uncertainty change for one stellar parameter at a time in Table \ref{tab:error}. 

The uncertainty for the temperature determination from the CO band-head indices are $\pm 150$\,K \citep{schultheis:16}. 

The uncertainties in the surface gravity, \logg, determined from isochrone fitting can be estimated by propagating the uncertainty in temperature through our procedure. For the typical giant in our sample, GC16763, this leads to \logg$=0.42$ (for \teff$=3450$\,K) and \logg$=0.91$ (for \teff$=3750$\,K). Including the uncertainties in the metallicity determination, we estimate the uncertainty in our \logg\ determination to $\Delta$\logg$= 0.3$\,dex.

The uncertainties in the microturbulence is difficult to assess, since it is an ad hoc parameter. The values are within $0.6$\,\kms\ of each other, according to the prescription we use. We therefore set the uncertainty to  $\Delta \xi_\mathrm{micro}= 0.3$\,\kms.

The uncertainty in [$\alpha$/Fe] is estimated to be well under 0.2 dex with the typical range $-0.1 <  [\alpha/Fe] < +0.4$. We see from Table \ref{tab:error} that 
the uncertainties in \logg, microturbulence, but also the [$\alpha$/Fe] abundance, all contribute at a $0.1-0.2$\,dex level. The uncertainties are certainly correlated. One way of assessing the total uncertainty is the perform our \logg\ determination for the spread in temperature. The two last rows in the Table presents these results. The uncertainty is within 0.2 dex.

The total uncertainty in [Fe/H] is thus of the order of $\pm 0.2$\,dex.

\begin{deluxetable}{l c c c}
\tablecaption{Uncertainties in the derived metallicities, [Fe/H], due to uncertainties in parameters for a typical giant in our sample (GC16763): \teff $=3600$\,K, \logg $=0.65$, $\mathrm{[Fe/H]}=-0.3$, $\xi_\mathrm{micro}=2.3$\,\kms, and [$\alpha$/Fe] $=0.2$. \label{tab:error}}
\tablewidth{0pt}
\tablehead{
\colhead{Parameter} & \colhead{change} &\colhead{$\Delta$[Fe/H]}   \\
   & \colhead{} & \colhead{(dex)} 
  } 
\startdata
$\Delta$\teff & $-150\,\mathrm{K}$  & $+0.04$  \\
              & $+150\,\mathrm{K}$  & $-0.02$  \\
$\Delta$\logg & $-0.3$\,dex & $-0.04$  \\
              & $+0.3$\,dex & $+0.19$\\
$\Delta$ $\xi_\mathrm{micro}$ & $-0.3$\,\kms &  $+0.11$\\ 
                      & $+0.3$\,\kms & $-0.10$\\                  
$\Delta$[$\alpha$/Fe] & $-0.2$\,dex & $-0.09$ \\
                      & $+0.2$\,dex & $+0.14$\\  
\multicolumn{2}{l}{3450\,K$\Big/$\logg$=0.42\Big/\xi_\mathrm{micro}=2.4$\,\kms} & $-0.03$\\
\multicolumn{2}{l}{3750\,K\Big/\logg$=0.91\Big/\xi_\mathrm{micro}=2.1$\,\kms} & $+0.17$\\
\enddata
\end{deluxetable}





\section{Results}

Figure \ref{fig:twospectra} and \ref{fig:spectra} show spectra of all of the 18 giants. It shows a portion of the observed wavelength region where four Fe lines are indicated. The  observations, shifted to the laboratory frame (with radial velocities, $v^\mathrm{LSR}_\mathrm{rad}$, values given in Table \ref{tab:kin}), are shown by the black underlying spectra and the synthetic spectra are overlaid in red. The stellar parameters are indicated above each spectrum. We have also plotted the telluric spectrum in green, shifted with the same velocity as the target star.  This shows how well the telluric division has performed. In other regions where the telluric contamination is more severe, this  overlaid telluric spectrum shows where to trust the spectrum the most and where spurious features can be explained by residuals from the division.

Our derived stellar parameters including the metallicity, \feh, are given in Table\,\ref{tab:phot}. Typical, estimated uncertainties in the derived metallicities are $\Delta$[Fe/H]$=0.2$\,dex.

\begin{figure}[!tbp]
  \centering
\epsscale{1.00}
\includegraphics[trim={0cm 0cm 0cm 0cm},clip,angle=90,width=1.00\hsize]{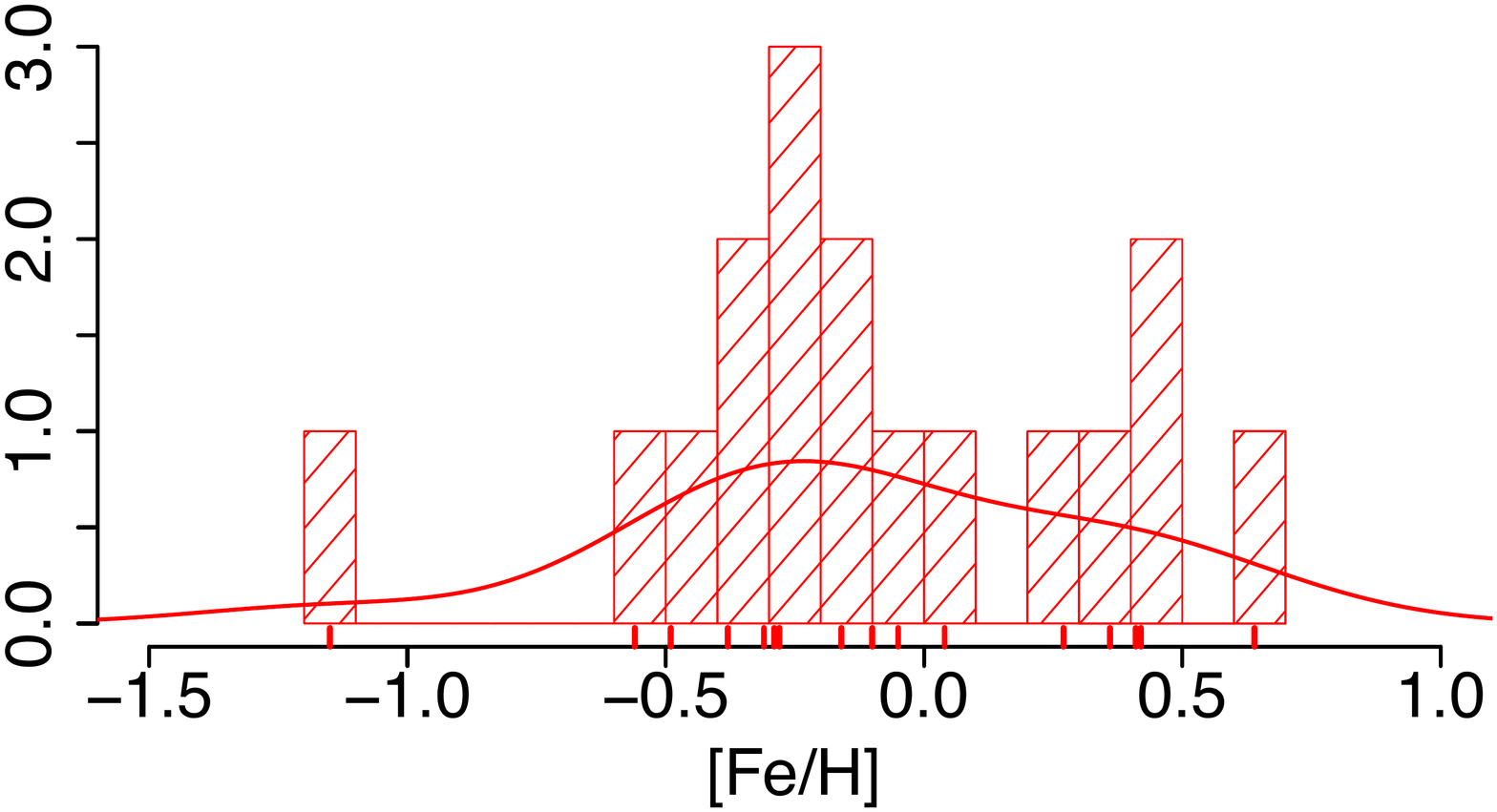} 
\caption{Histogram of the derived metallicities of stars in the Nuclear Star Cluster. The actual metallicity values are shown by the tick red marks below the histogram. The full line is a KDE (kernel density estimation) using Silverman's rule-of-thumb bandwidth \citep{feigelson}\ of the metallicities, shown to visualize the data. Here we have excluded the star GC7688 since we will find that it probably is not a star in the Nuclear Star Cluster, see Section \ref{membership} \label{fig:hist_new}.}
\end{figure}

\begin{figure}[!tbp]
  \centering
\epsscale{1.00}
\includegraphics[trim={0cm 0cm 0cm 0cm},clip,angle=90,width=1.00\hsize]{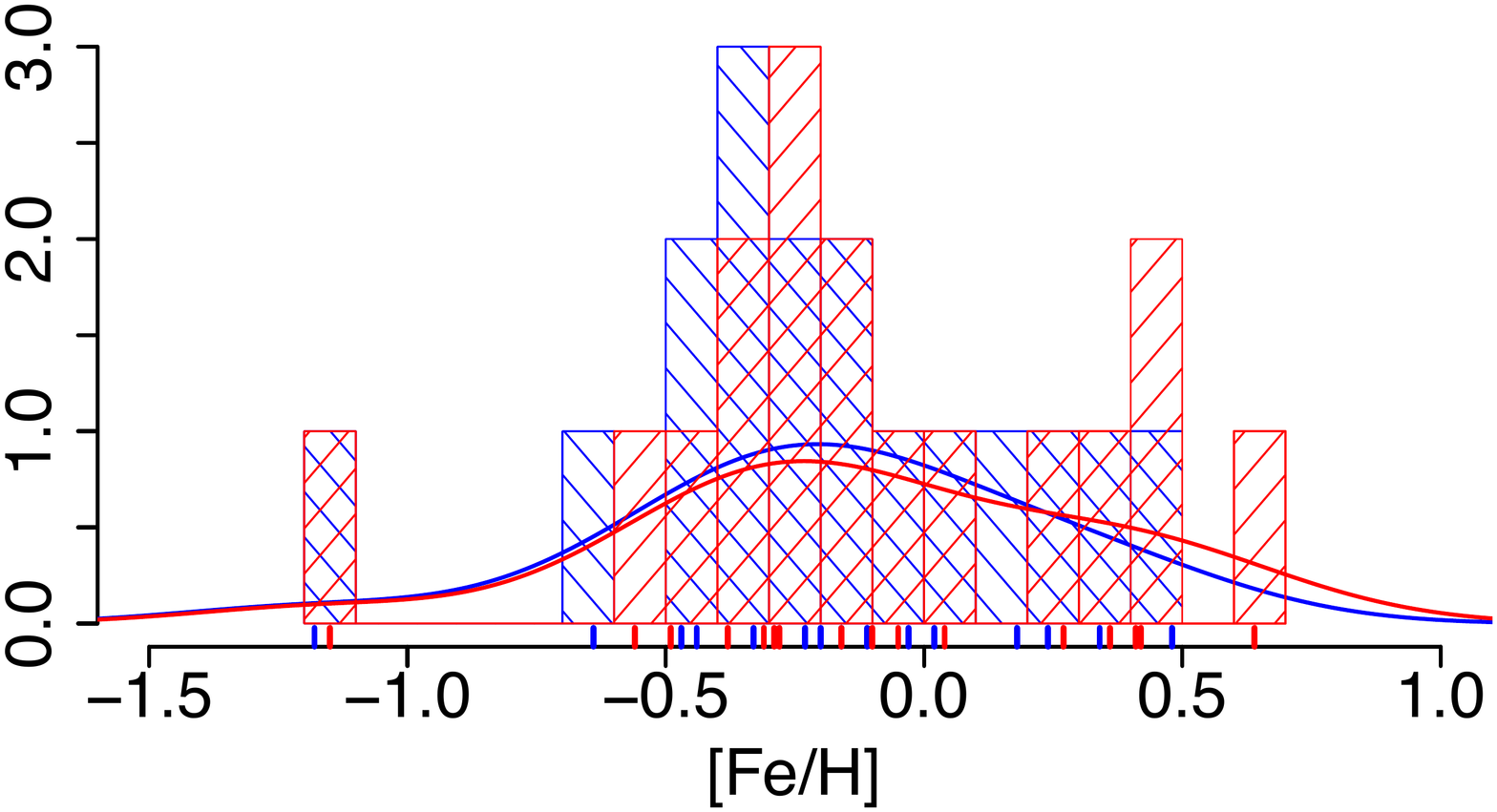} 
\caption{Histogram in blue of metallicities of stars in the Nuclear Star Cluster, derived using photometrically determined \logg. The red histogram is identical to that presented in Figure \ref{fig:hist_new}. The full lines are KDEs of the metallicities to guide the eye.  \label{fig:hist_phot}}
\end{figure}

The  distributions of our derived metallicities are shown in Figures \ref{fig:hist_new} and \ref{fig:hist_phot}, both as histograms and as  Kernel density estimations, smoothed with a Gaussian kernel using Silverman's rule-of-thumb bandwidth \citep{feigelson}. The latter is preferred over histograms due to the problems with the choice of bin size for a histogram and choice of bin positions \citep[see][]{feigelson}. Figure \ref{fig:hist_new} shows the metallicity distribution for the analysis in which the surface gravities are derived from placing the star on  isochrones, given the effective temperatures and metallicities of the stars (see Sect. \ref{logg}). 
Figure  \ref{fig:hist_phot} shows the metallicity distribution for the analysis  in which the surface gravities are determined from photometry. 


\section{Discussion}


\subsection{General Remarks}
Our goal is to determine the distribution of metallicities of the stars in the Nuclear Star Cluster of the Milky Way with the ultimate goal to explore the trend of [$\alpha$/Fe] vs [Fe/H]. In order to achieve this goal, with as high precision and accuracy as possible, high-resolution spectra ($R\gtrsim 20,000$) with good signal-to-noise ratios are needed.  The decreased sensitivity to the metallicity, severe atomic and molecular blending of cool stars, and increased difficulties to deal with telluric lines can otherwise be a large source of uncertainty. 
At low spectral resolution, when neighboring lines start to blend,  unknown or missing blends are likely to compromise the abundance determination, not to mention the additional strong line issue. 
In wavelength regions with severe telluric contamination, as high spectral resolution as possible is needed to disentangle the telluric spectrum.

Furthermore, the target stars have to be observed in the K-band ($2.2-2.4$\,\mic) due to the extreme optical extinction toward the center of the Galaxy. However, the strong telluric lines represent a complicating factor.  This means that new methodology needs to be developed since traditional ones using optical spectroscopy and photometry are unusable. High resolution spectroscopy in the K band have been used infrequently for abundance analyses, which means that properties of the spectral lines in this wavelength region are not as well studied as, for example, in the optical region. We have developed our K-band line list (Thorsbro et al. 2018) that is the source of our Fe lines used in this analysis. Similarly, the APOGEE project \citep{apogee_ref} had to develop their own proper line list to be able to perform their own  spectral analysis of H-band spectra \citep{smith:13,apogee:15}.

Spectroscopic observations of Galactic Center stars are challenging, especially for metal-rich, cool giants. Compromising with spectral resolution and/or the signal-to-noise level is tempting and yields a larger number of stars that can be investigated, but will inevitably lead to large uncertainties. In these type of spectra the abundance analysis tends to be biased toward the strongest lines in the spectrum, lines that should not be used for an abundance determination because they are quite sensitive to the microturbulence and only weakly sensitive to abundance, the parameter we wish to measure. In the worst case the abundance is totally drowned in the uncertainty of the microturbulence. These lines are often saturated and lie on the flat part of the curve-of-growth \citep[see for example][]{gray:2005}. A change in the microturbulence of $0.5$\,\kms\ for, e.g., a weak Sc line with an equivalent width of $W=50$\,m\AA\ leads to a unnoticeable change in the abundance of $\delta$[Sc/Fe]$=0.01$, whereas a saturated line with $W=350$\,m\AA\ would change by as much as 0.4 dex. Weak lines are therefore the preferred diagnostics for the determination of elemental abundances. It should, at the same time, be noted that the lower the S/N, the more uncertain these weak lines' abundance will get. That is why spectra of high resolution {\it and} high S/N are necessary for an precise abundance determination.

In the K band, lines in a typical M giant start to saturate at an equivalent width of $W\sim300\,$m\AA, which corresponds to a typical line-depth in the core of lines of 0.75 of the continuum, at a resolution of $R=24,000$. A very strong line, with small pressure broadening, reaches down to 0.55 of the continuum. Lines stronger than this should be avoided. A massively strong line can reach 0.35, and no line can get deeper than this in the near-IR if formed in LTE. Note that this is different for lines formed in the ultraviolet. In that wavelength region the lines  can get much deeper, since the gradient of the Planck function with respect to the temperature in the atmosphere is greater. 

If an infrared spectral line is not formed in LTE and scattering plays a large role for the line source function, the line might be expected to be stronger. This may be the reason for the strong Sc lines at $22395$\,\AA\, (see Figures \ref{fig:twospectra} and \ref{fig:spectra}), which are much stronger than any reasonable Sc abundance can result in, for a line formed in LTE. This type of line should obviously not be used to measure the over-all metallicity of a star at any spectral resolution and will disturb global spectral-fitting routines in LTE.



We have chosen the spectral resolution and signal-to-noise ratios (or exposure times) such that we can obtain spectra of such high quality that we can optimize for a detailed abundance analysis while keeping the uncertainties to a minimum.  


\subsection{Membership of our targets of the Nuclear Star Cluster \label{membership}}


\begin{deluxetable*}{l c c c c l  }
\tablecaption{Spectroscopic stellar parameters and the probable locations of the stars in the nuclear components (Nuclear Star Cluster (NSC), Nuclear Disk (ND), or foreground Bulge (FB) \label{tab:member}}
\tablewidth{0pt}
\tablehead{
\colhead{Star} &  \colhead{$T_{\mathrm{eff}}$} & \colhead{$\log g$} &  \colhead{\feh} & \colhead{$\xi_{\mathrm{mic}}$} &  \colhead{Location} \\
\colhead{} & \colhead{[K]} & \colhead{(dex)} & \colhead{(dex)} & \colhead{[\kms]} &  \colhead{}   } 
\startdata
GC10812 &  3800& 0.34 &  $-1.15$  & 2.5  & NSC  \\
GC11473 &  3550& 1.13 &  $+0.64$  & 2.0  & ND   \\
GC13282 &  3700& 0.97 &  $-0.05$  & 2.1  & NSC   \\
GC14024 &  3650& 1.23 &  $+0.42$  & 2.0  & NSC   \\
GC16887 &  3500& 0.96 &  $+0.41$  & 2.1  & NSC   \\
GC6227  &  3800& 0.95 &  $-0.38$  & 2.1  & NSC   \\
GC7688  &  4150& 1.78 &  $-0.08$  & 1.8  & FB  \\
GC15540 &  3600& 0.72 &  $-0.16$  & 2.2  & NSC   \\
GC10195 &  3500& 0.68 &  $+0.04$  & 2.3  & NSC \\
GC11532 &  3500& 0.35 &  $-0.56$  & 2.5  & NSC  \\
GC13882 &  3300& 0.15 &  $-0.31$  & 2.6  & NSC \\
GC14724 &  3600& 0.55 &  $-0.49$  & 2.3  & NSC   \\
GC16763 &  3600& 0.65 &  $-0.29$  & 2.3  & NSC  \\
GC7104  &  3750& 0.92 &  $-0.28$  & 2.1  & ND \\
GC11025 &  3400& 0.69 &  $+0.27$  & 2.3  & NSC \\
GC16867 &  3650& 0.74 &  $-0.29$  & 2.2  & NSC \\
GC16895 &  3450& 0.51 &  $-0.10$  & 2.4  & NSC    \\
GC8631  &  3900& 1.63 &  $+0.36$  & 1.9  & NSC  \\
\enddata
\end{deluxetable*}

\begin{figure}
  \centering
\epsscale{1.00}
\includegraphics[trim={0cm 0cm 0cm 0cm},clip,angle=-90,width=1.00\hsize]{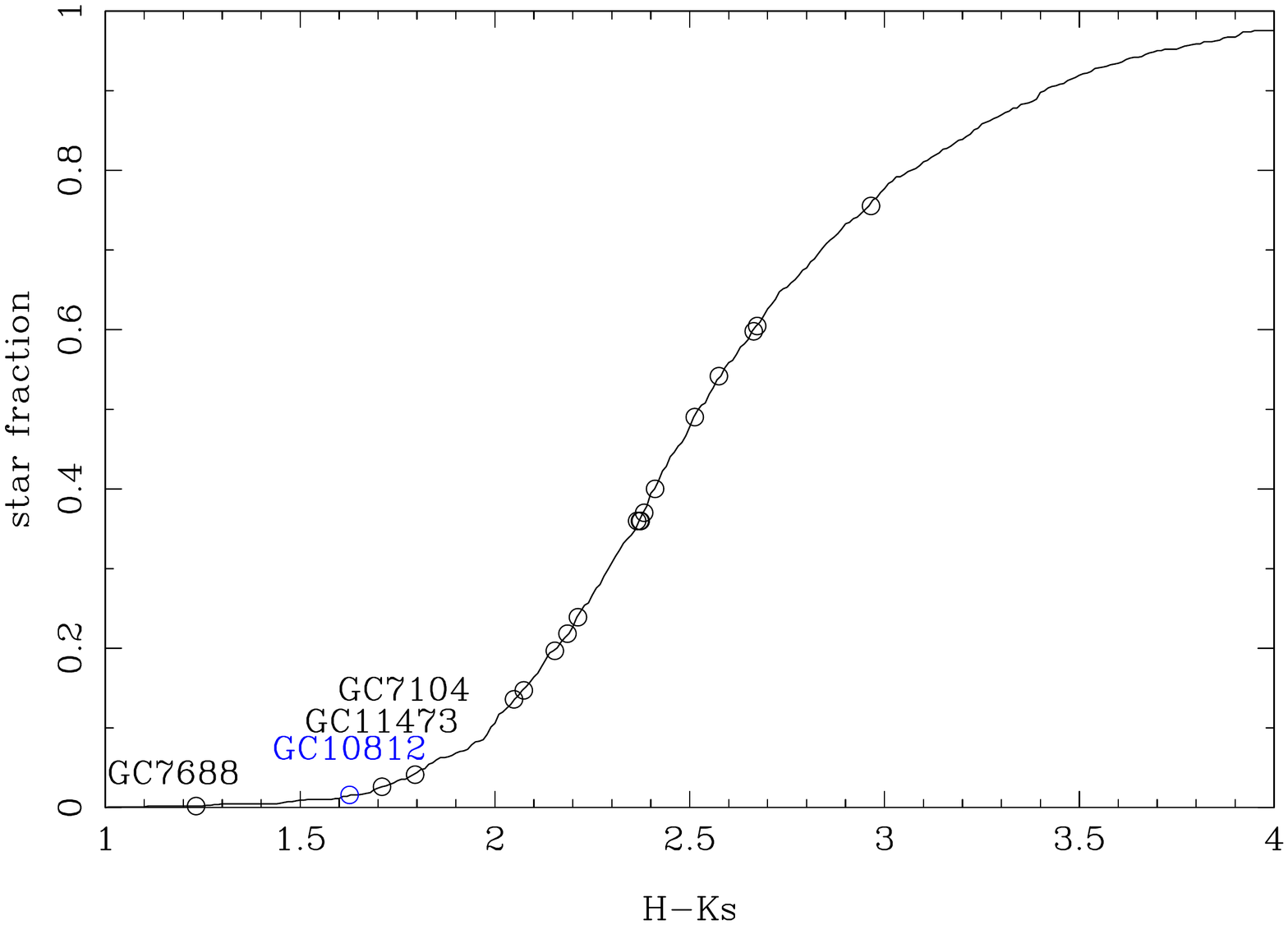} 
\caption{A cumulative plot of $H-K_s$ colors of stars in the Galactic Center region. Our targets are marked with circles and the four bluest stars are indicated. The metal-poor giant, GC10812, discussed in \citet{ryde:16} is indicated in blue. \label{fig:fritz1}}
\end{figure}

Direct distance measurements cannot determine whether the stars are in the nuclear region, since the nuclear region has a radius of about $230$\,pc \citep{launhardt02}, which is 
small compared to the distance to the Galactic Center of about 8.3 kpc \citep{GCdistance}. Thus, a direct distance precision of less than 2\% would be necessary to be useful. That is far from possible for giants. There is, however, high extinction within the Galactic Center \citep{chatzopoulos15b}, a  fact can be used to determine distances relative to the very center. 

The concept, which is very similar to \citet{Ryde:GC:MPG}, is the following: We determine the fraction of the comparison stars ($f$) bluer than the target star and then use the nuclear component model of \citet{GCdistance} to determine at which distance, $z$, from the very center the star would need to be placed, so that $f$ of all stars in the model are in front of $z$.
The three-dimensional model of \citet{chatzopoulos15b}, which we use, is obtained by fitting Galactic Center data, accounting for flattening. Although it does not include bulge-bar and disk components, it is preferable for stars close to the center, since the nuclear components are not modeled at all in other models. Combining the model of \citet{chatzopoulos15b} with bulge models does not work, because in the latter the nuclear region is included in the bulge fit.

The comparison stars are stars with an extinction corrected ${K_s}<9$ from \citet{nishiyama2009}. With this magnitude limit, we avoid that very red stars are too faint to be in the sample. Usually we use stars within $100\arcsec$ of Sgr~A* to limit the influence of extinction variations in the plane of the sky, but we also use a second sample out to $500\arcsec$ for very blue stars because the other sample contains too few very blue stars. We add 0.22 in quadrature to the general uncertainties of the measured color $H-K_s$. This number follows from the variation of $A_{Ks}$ extinction in the plane of sky \citep{schoedel10} and the extinction law of \citet{fritz:11} which we use. Due to the patchy nature of the extinction, larger uncertainties are not excluded. We do not account for intrinsic color variations, as these are small for luminous giants in $H-K_s$ compared to the extinction variation. 

\begin{figure}
  \centering
\epsscale{1.00}
\includegraphics[trim={0cm 0cm 0cm 0cm},clip,angle=-90,width=1.00\hsize]{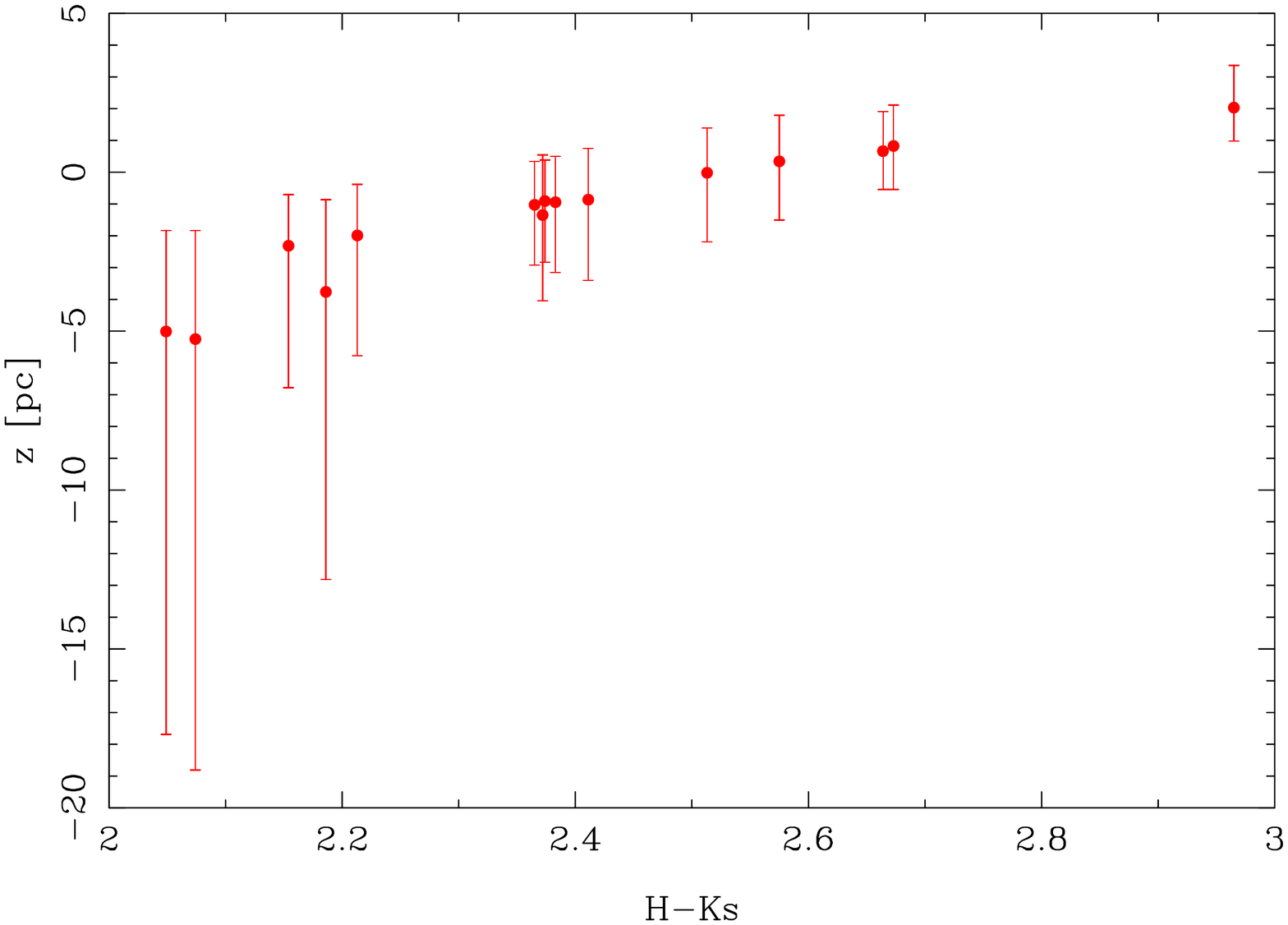} 
\caption{From the extinction derived distance of the targets relative to the Galactic Center. Negative z are in front of the center. Three stars at small H-K and z are off the scales of this plot, their z are given in the text.  \label{fig:fritz2}}
\end{figure}

In Figure \ref{fig:fritz1}, we compare the colors of our targets with the mentioned comparison sample of \citet{nishiyama2009}. Most stars have typical colors. Four stars are bluer than 90\% of the stars, which therefore might place them in the foreground.  Two of these stars (GC11473 and GC7104), are redder than GC10812, for which we made a detailed orbit calculation in \citet{Ryde:GC:MPG} placing it in the Nuclear Disk or Cluster. These two stars are thus likely to be located there. We note though that GC10812 is of lower metallicity, \feh$<-1.0$, than the other stars in our sample.  The most likely distances from the Galactic Center of these two stars are (including 1-sigma limits):  $D(\mathrm{GC11473})=-36_{-109}^{+29}$\,pc and $D(\mathrm{GC7104})=-12_{-84}^{+8}$\,pc, thus putting them well within the Nuclear Disk. In Figure \ref{fig:fritz2} we give the most probable distances for the other 15 stars of the sample. All these stars are redder than the discussed stars and are possible or likely members of the Nuclear Star Cluster, which has a half-light radius of about 4 to 9\,pc \citep{fritz16,schoedel14}. 

GC7688 is by far the bluest star in our sample but is not metal rich and appears to have high \teff\ (Table~\ref{tab:phot}). Its distance to the Galactic Center, estimated by our model, is $D(\mathrm{GC7688})=-282_{-241}^{+125}$\,pc. While this range overlaps with the Nuclear Disk size of \citet{launhardt02}, we think that for this star the omission of the bulge and disk in our model, leads to a relevant distance underestimate. Therefore, the star is probably a bulge member.

We now examine the dynamics of our stars in order to test their membership in the nuclear region during their entire orbit. For the kinematics, we mainly use the radial velocities published in \citet{fritz16}. For most of our stars, proper motions were obtained in that effort, but were not used in that analysis and catalog because their precision was too low to derive reliable masses from the dispersions. For our purpose, however, of comparing a few stars with a known population, a lower precision is sufficient.  We use all stars with an uncertainty $<2$~mas\,yr$^{-1}$. For four stars, we obtain new proper motions using data obtained in the ESO programs 087-B-0182 and 091-B-0172, providing us with a baseline of about 2 years. The proper motions are obtained in a similar way as in \citet{fritz:11}; the distortion is corrected by applying the radially symmetric formalism first used in \citet{Trippe_08}. It was shown in \citet{Fritz_10} that the distortion is stable between instrument interventions during the epochs of observations, thus it is acceptable that the adopted parameters for the distortion correction were obtained in other observations within the same intervention cycle. The uncertainty consists of the scatter between different images, and of, as a dominating component, a conservative uncertainty floor of 1 mas per epoch. 
For a few of our stars, it was not possible to derive proper motions, because too few good images were available. 

\begin{figure}
  \centering
\epsscale{1.00}
\includegraphics[trim={0cm 0cm 0cm 0cm},clip,angle=-90,width=1.00\hsize]{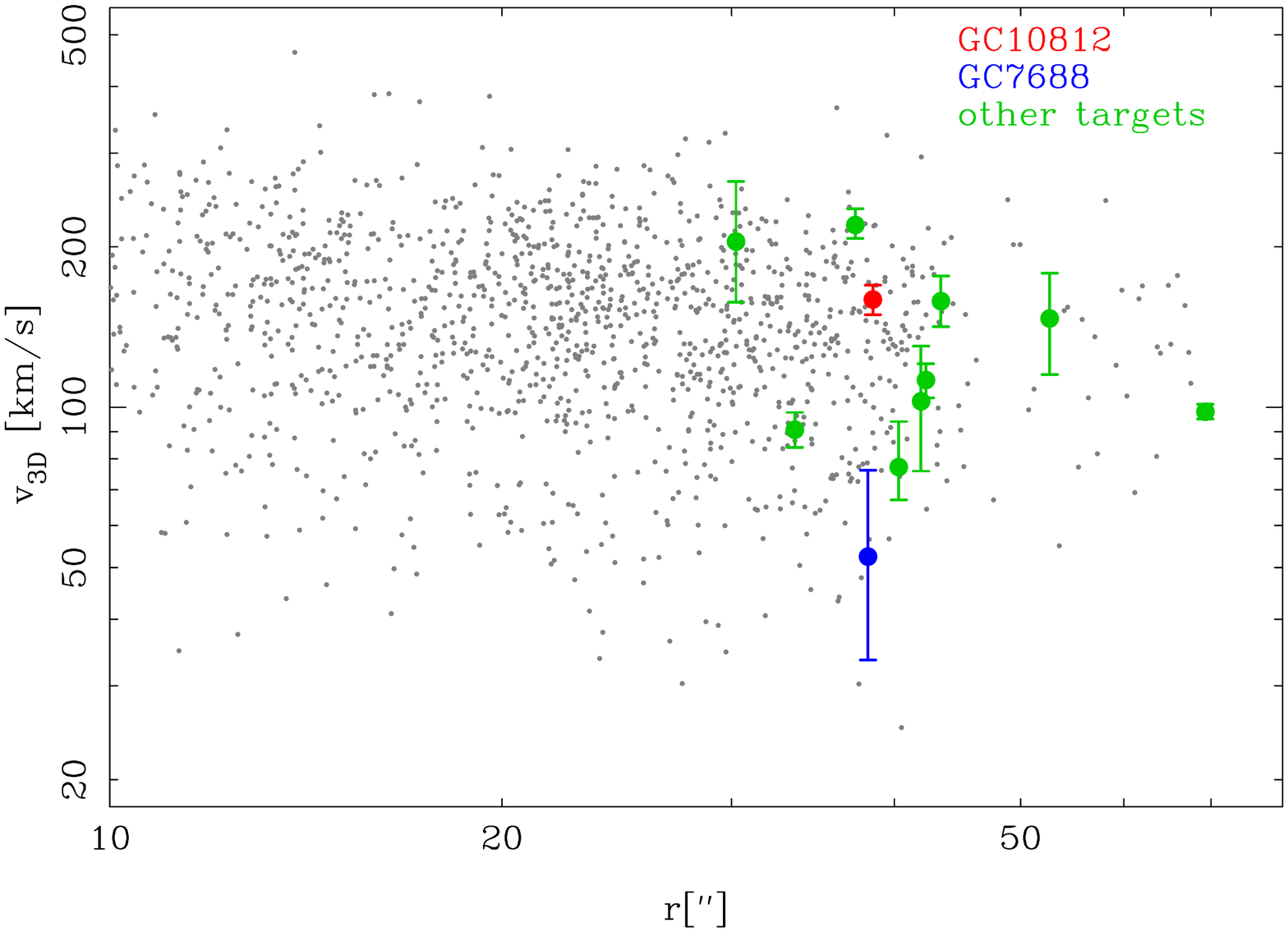} 
\includegraphics[trim={0cm 0cm 0cm 0cm},clip,angle=-90,width=1.00\hsize]{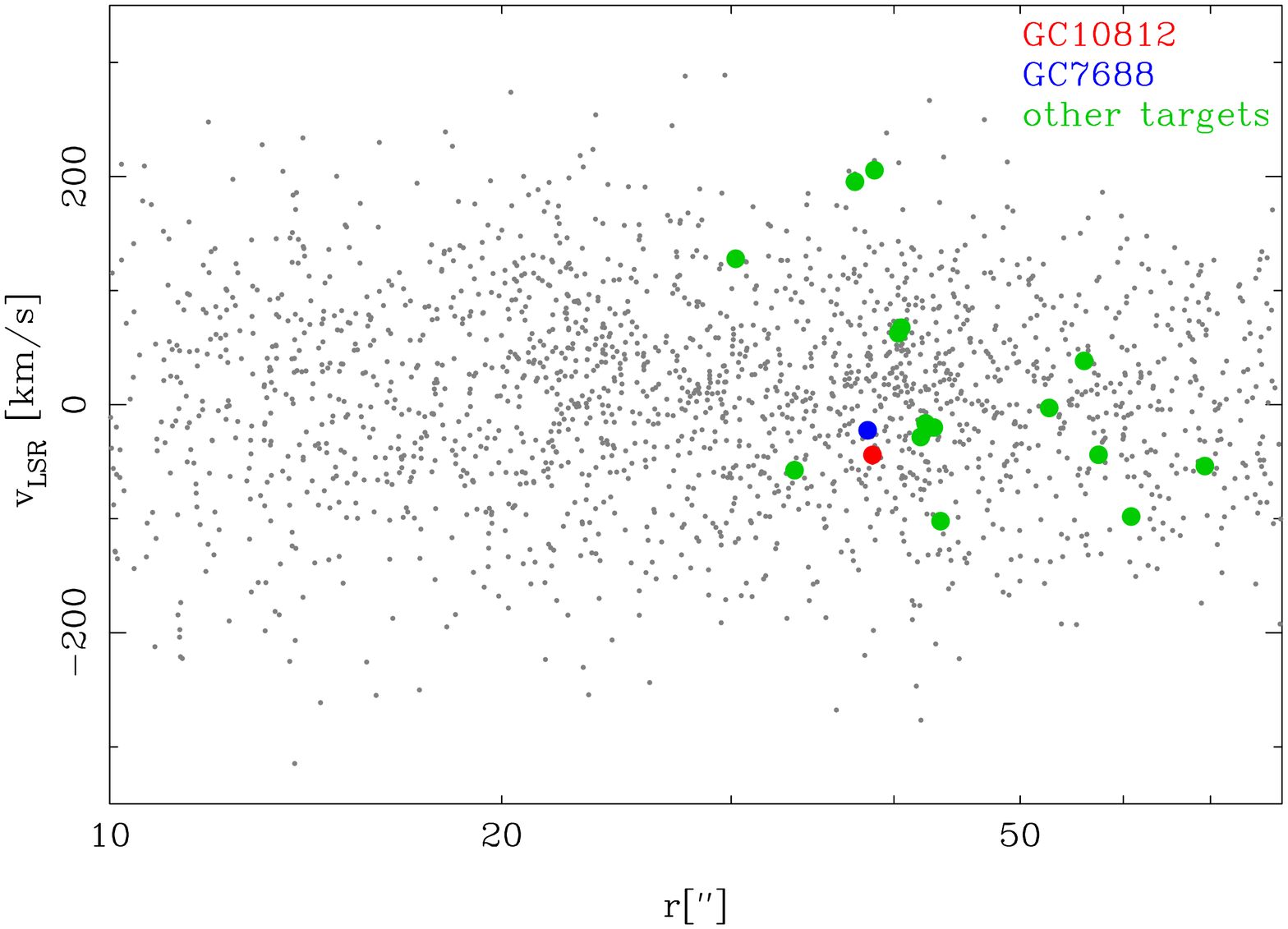} 
\includegraphics[trim={0cm 0cm 0cm 0cm},clip,angle=-90,width=1.00\hsize]{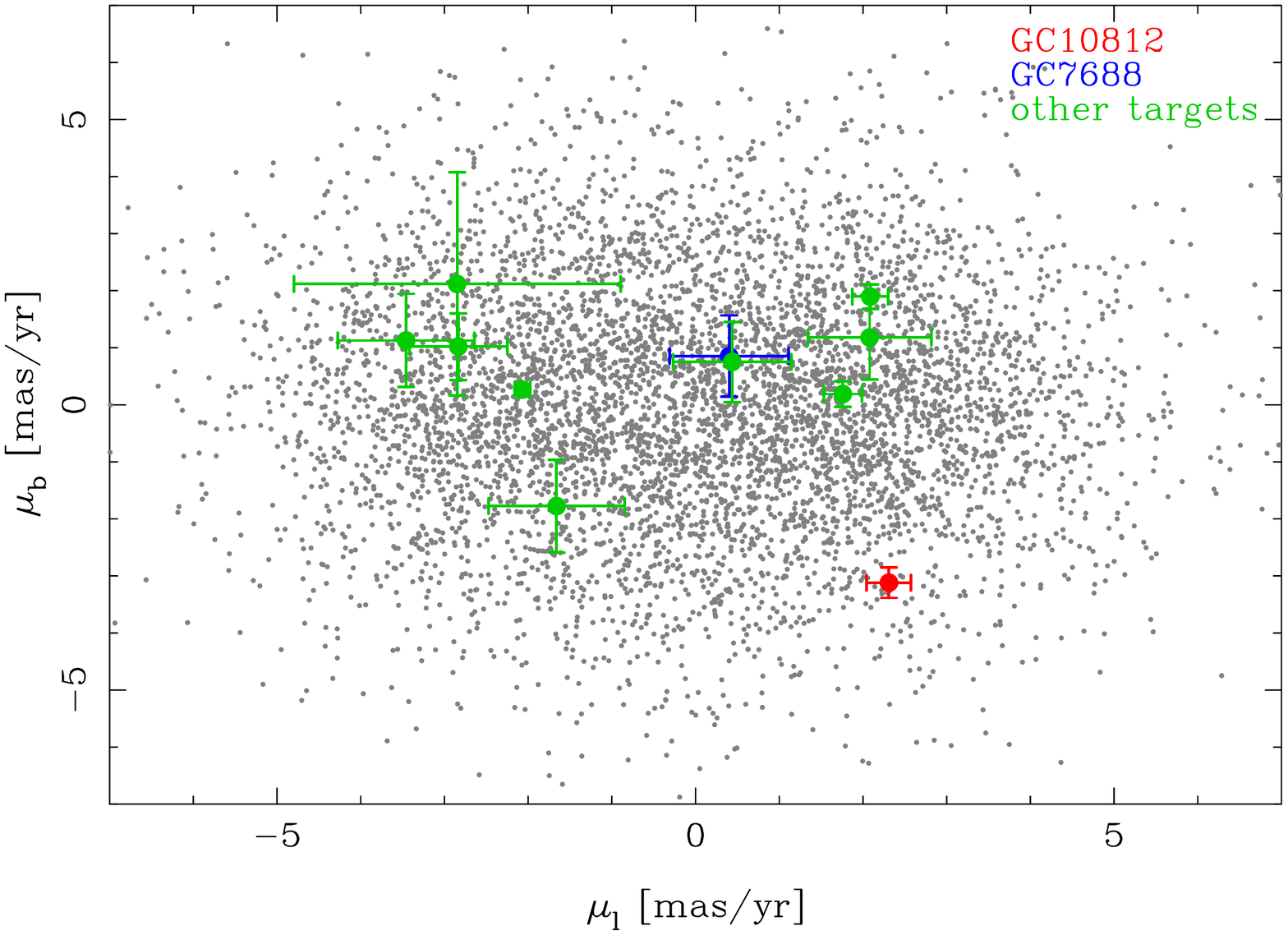} 
\caption{Kinematics of the targets. The plots show the space velocity velocity (upper panel), line of sight velocities (middle panel), and proper motions of the targets (colored; lower panel) in comparison with other stars in the Galactic Center (gray). 
In the top panel we assume a distance of 8.3 kpc for the conversion of proper motions to velocities. Since some stars do not have a proper motion measurement, some stars are not shown in the space velocity and proper motion plots. In the bottom plot we only show only comparison stars which have $r>20\arcsec$, since the velocities of stars are larger in the center and our stars are also outside the central 20$\arcsec$. \label{fig:kinematics}}
\end{figure}

Our radial velocities are measured from NIRSPEC, since they have a better precision than the SINFONI velocities. Considering their uncertainties,  the two are consistent. 
The uncertainty in the velocities is about $1\,$km\,s$^{-1}$ with about equal contributions from wavelength calibration, SNR and LSR uncertainties. All velocities are listed in Table~\ref{tab:kin} and are shown in Figure~\ref{fig:kinematics}.

We find that the motions of our stars are similar to typical stars in the region. Thus, it strengthens our conclusions that most stars are members of a nuclear component. In particular, the relative blue star GC7104 has a 3D motion,  which is smaller than the 3D motion of star GC10812. Since we have shown in \citet{ryde:16} that the star GC10812 does not leave the nuclear environment,  GC7104 is certainly a nuclear star.
The other medium-blue star (GC11473) has, unfortunately, no proper motions determined and a  $v_\mathrm{rad}^\mathrm{LSR}=206$~km\,s$^{-1}$. 
This star could be member of a nuclear component, since the other stars have similar velocities. Since most of the other stars stay in the nucleus, this star probably stays there as well. However, we cannot be certain without determined proper motions. The bluest star GC7688, has one of the smallest space velocities of all stars. Its proper motion is within nearly 1 $\sigma$ of zero and $v_\mathrm{rad}^\mathrm{LSR}=-22$ km\,s$^{-1}$ is rather small. From that follows that the star is currently close to its apocenter and we conclude it is like that over most of its eccentric orbit; it will be closer to the center compared to its current location. Nevertheless, even if it enters into the nuclear region, it is still clearly different from the typical nuclear stars since most are not on such eccentric orbits. 

Overall, we find that all but three of our target stars are likely or possibly members of the Nuclear Cluster, see Table \ref{tab:member}. From the remaining three, one (GC7104) is nearly certain a member of the Nuclear Disk and GC11473 is probably also a member. One star, GC7688, is likely not a member of a nuclear component. We have therefore excluded this star from the the sample of NSC giants, and omit it from the histogram in Figure \ref{fig:hist_new} and \ref{fig:hist_phot}.


\subsection{Metallicity Distribution}

The distributions in Figures \ref{fig:hist_new} and \ref{fig:hist_phot} actually only show the 17 stars located in the Nuclear Star Cluster. We derive a metallicity distribution that is broad, 
spanning $~-0.5 <$\feh$< +0.5$  (see Fig.~\ref{fig:hist_new}).  In finding no stars with \feh$>+0.6$, the extremely high metallicities of \citet{do:15} would not appear to be confirmed, and  we find very few stars with \feh$<-0.7$ 
(less than $\sim6\%$). 
We recall that the uncertainties in the metallicities of the order of 0.2 dex. Also, the width and form of the metallicity distribution is not significantly different between the distributions based on the two different methods for the determination of the surface gravity, \logg.

Our determined velocities and derived proper motions of our stars are give in Table~\ref{tab:kin}. In order to determine the velocity dispersion, we divided our sample of Nuclear Cluster stars (i.e. excluding GC7688) in two groups in metallicity with a division at solar metallicity. Using only radial velocities (which we have for all 17 stars) we get $\sigma=94\pm24\,$\kms\ for the metal-poor group and $\sigma=87\pm20\,$\kms\ for the metal-rich group. 
If we determine the velocity dispersion using also the proper motions, and treating all dimensions the same, we arrive at $\sigma=79\pm14\,$\kms\ and $\sigma=111\pm18\,$\kms, respectively.
Thus, we see no kinematic difference between the groups. Our stars are, on average, about $41\,\arcsec$ from Sgr~* in projection, where the dispersion is about $83\,$\kms. Our values are thus consistent with this value.

At first glance, our abundance distribution appears similar to that of the Galactic bulge/bar and most significantly, is not populated with stars at the extremes of the abundance distribution, nor did we exclude from our sample any such candidates.   It is also noteworthy that the abundance distribution is inconsistent with being a Solar metallicity disk population or consistent with an inward extension of the metal poor halo.



Our metallicity distribution  based on high-resolution spectroscopy of the old population, can be contrasted with high-resolution studies by \citet{carr:00,ramirez:00a,cunha:2007:apj,davies:09} probing the current-day compositions by observing young ($<1~$Gyr), luminous cool cluster giants within 30 pc of the Galactic Center. They find a narrow, near-solar iron abundance distribution.  Compared to these earlier efforts, our distribution is more broad, and indeed other studies that address the fainter population of red giants are finding broader and more complex abundance distributions.  Earlier work by Ryde \& Schultheis (2015) suggested that the alpha abundances are also similar to the Galactic bulge; we may therefore acknowledge a growing picture in which the central cluster includes an older $>$ few, perhaps 10 Gyr population that resembles the bulge/bar more than other stellar populations.  Based on existing sample sizes, it is quite premature to even discuss chemical evolution models.  However, it will be valuable to do careful differential measurements between ever fainter giants and the "manifest" supergiant population- such differential measurements may be the most valuable as we consider how the present day starburst differs in chemistry from the foundational old stellar population.



\subsection{How metal-rich are the metal-rich stars?}

The Galactic center is a unique region where one might expect to find the most metal-rich stars in the Galaxy. The reason is due to  the presumed early, strong burst of star-formation with high star-formation efficiency that this region must have experienced according to chemical-evolution models \citep[see e.g.][]{grieco:15}. The question is then if and how many super metal rich stars can be formed. One might worry about the spectral analysis of these very metal-rich stars since lines get increasingly saturated and therefore less abundance sensitive.

We have derived higher-than normal metallicities (\feh $> +0.5$\,dex) for one star in our sample. Even though other authors have also found stars of such high metallicities (see, for example, APOGEE \citep{apogee_ref}, Gaia-ESO \citep{rojas-arriagada:17}, and  \citet{do:15}), the difficulties in an abundance analysis increases for higher metallicities, with line crowding and lines getting stronger and more saturated, which means an increased sensitivity to microturbulence. The question is how much we can trust the highest metallicities. 

In Table \ref{tab:max} we present the metallicities derived for the metal-rich star GC11473 for a few models with varying stellar parameters. The two first rows give the metallicities derived from models with  surface gravities based on isochrones and on photometry, respectively.  The high metallicity derived is very sensitive to the \logg\ and microturbulence. In order to investigate what the lower uncertainty limit is for the metallicity determination, we decrease the \teff\ within the uncertainty range and change the \logg\ (based on isochrones, see Section \ref{logg}) and the microturbulence, accordingly. We then derive a metallicity of \feh=0.4. Furthermore, allowing for a reasonable increase of the microturbulence of 0.3\,\kms, the metallicity will decrease an additional 0.1 dex. Thus, due to the uncertainties affecting the most metal-rich stars, our highest metallicity star has such uncertainties that a more normal metallicity of \feh$=+0.3$ to $+0.4$ dex is still consistent with our observed spectrum.   

\begin{deluxetable}{l c c}
\tablecaption{Uncertainties in the derived metallicities, [Fe/H], of our most metal-rich star (GC11473): \teff $=3550$\,K, \logg $=1.13$, $\mathrm{[Fe/H]}=+0.6$, $\xi_\mathrm{micro}=2.0$\,\kms, and [$\alpha$/Fe] $=0.0$. \label{tab:max}}
\tablewidth{0pt}
\tablehead{
\colhead{Model} & \colhead{parameters}  &\colhead{[Fe/H]}   \\
   & \colhead{\teff$\Big/$ \logg$\Big/$ $\xi_\mathrm{micro}$} &  \\
   & \colhead{[K]$\Big/$(dex)$\Big/$[\kms]} & \colhead{(dex)}    
  } 
\startdata
isochrone & $3550\Big/1.13\Big/2.0$ & $0.6$ \\ 
photometric & $3550\Big/0.97\Big/2.1$ & $0.4$ \\ 
isochrone: & & \\ 
$\Delta$\teff$=-150$\,K & $3400\Big/0.84\Big/2.2$ & $0.4$\\ 
isochrone: & & \\ 
$\Delta$\teff$-150$\,K; & & \\
$\Delta\xi=0.3$\,\kms   &$3400\Big/0.84\Big/2.5$ & $0.3$\\ 
\enddata
\end{deluxetable}







\section{Conclusions}

We have analyzed high signal-to-noise and high-resolution K-band spectra of 18 M giants with $10.5 <K_s< 12$, projected $\sim 1-2$\,pc near the Galactic Center.  These are among the faintest Galactic center giants studied at high resolution, and their stellar parameters place them as intermediate-age to old red giants.  This work provides a window into the nature of what is likely the ancient foundational population of the Galactic nuclear cluster.  Using additional 
proper motion measurements and radial velocities, we conclude that 17 of them are members of the nuclear cluster.  We find a broad metallicity distribution ranging from \feh$\sim-0.5$ to \feh$\sim+0.5$\,dex and no evidence that the kinematics depends on [Fe/H].  We find no additional stars near [Fe/H]=$-1$, and even the most metal rich stars in our sample can be analyzed and fall within the range of +0.5 dex, the upper limit for the bulge as found by previous studies. The most metal rich star, GC1143, is found to have \feh$=+0.64$. However, uncertainties in the stellar parameters, might in principle bring this down to \feh$\sim 0.3-0.4$\,dex. 
Given our small sample, we can state that the MDF is broadly similar to other fields in the Galactic bulge, but appears not to be as narrow as that found for the supergiants \citep{cunha:2007:apj}. A larger sample is necessary to explore hints of substructure in the abundance distribution and to confirm whether any correlation between abundance and kinematics is present.  If the substructure in the [Fe/H] distribution strengthens with increasing sample size, it may either reflect a globular cluster-like origin for the central cluster (similar to Terzan 5 and its complex abundance distribution e.g. Origlia et al. 2011) or multiple populations of a different origin.   Our next steps will be to explore the alpha elements in these stars, noting that \citet{cunha:2007:apj} found some enhanced alpha elements in the supergiant population, while \citet{ryde_schultheis:15} found scaled Solar abundances at the metal rich end. As the nuclear star cluster is likely to be a superposition of stellar populations of different ages and likely, enrichment histories, it will be of great importance to measure the trends of [$\alpha$/Fe] vs. [Fe/H], which may shed light on the history of the nuclear cluster.    




\acknowledgments
R.M.R. acknowledges support from grants AST-1413755 and AST-1518271 from the National Science Foundation.
N.R. acknowledges support from the Swedish Research
Council, VR (project number 621-2014-5640), Funds from Kungl. Fysiografiska
Sällskapet i Lund. (Stiftelsen Walter Gyllenbergs fond and Märta och
Erik Holmbergs donation), and from the project grant “The New Milky” from the Knut and Alice Wallenberg foundation. M.S. acknowledges the Programme National de Cosmologie et Galaxies (PNCG) of CNRS/INSU, France, for financial support. The authors wish to
recognize and acknowledge the very significant cultural role and 
reverence that the summit of Mauna Kea has always had within the
indigenous Hawaiian community.  We are most fortunate to have the
opportunity to conduct observations from this mountain.

{\it Facilities:} {KECK:II (NIRSPEC)}


\bibliographystyle{yahapj}

\end{document}